\PassOptionsToPackage{prologue,table}{xcolor}
\documentclass[acmtog]{acmart} %

\acmSubmissionID{595}

\gdef\cropInsets{0}

\usepackage{animate}

\usepackage{xargs}
\usepackage[prologue,table]{xcolor}
\usepackage{booktabs,makecell} %

\usepackage{tabularx}
\usepackage{caption}
\usepackage{nicematrix}
\NiceMatrixOptions{cell-space-limits=1.5pt}
\usepackage{float}
\usepackage{mathtools}
\usepackage{amsmath,amsfonts}
\usepackage{bm}
\usepackage{microtype}
\usepackage{units}
\usepackage{subcaption}
\usepackage{overpic}
\usepackage{cancel}
\usepackage{wrapfig}
\usepackage{placeins}
\usepackage{empheq}
\usepackage[export]{adjustbox}
\usepackage[vlined,boxed,ruled]{algorithm2e} %

\SetKw{Or}{or}
\SetKw{And}{and}
\SetStartEndCondition{ }{}{}%
\SetKwProg{Fn}{def}{\string:}{}
\SetKwFunction{Range}{range}%
\SetKw{KwTo}{in}
\SetKwFor{For}{for}{\string:}{}%
\SetKwFor{While}{while}{:}{fintq}%
\SetKwIF{If}{ElseIf}{Else}{if}{:}{elif}{else:}{}%
\SetKwComment{Comment}{\# }{}
\SetKw{Continue}{continue}
\AlgoDontDisplayBlockMarkers
\SetAlgoNoEnd
\SetAlgoNoLine
\DontPrintSemicolon

\SetAlFnt{\small}
\SetAlCapFnt{\small}
\SetAlCapNameFnt{\small}
\SetAlCapHSkip{0pt}
\IncMargin{-\parindent}

\usepackage{xr}
\usepackage{multirow}
\usepackage[capitalize,noabbrev]{cleveref} %
\usepackage{url} %
\usepackage[utf8]{inputenc}

\def\commentType{0}

\ifnum\commentType=0
    \newcommandx{\customComment}[3]{}
    \newcommandx{\customTODO}[3]{}
\fi
\ifnum\commentType=1
    \usepackage[
    type=upperleft,
    unit=in
    ]{fgruler}
    \newcommandx{\customComment}[3]{\textcolor{#2}{\textsl{#1: #3}}}
    \newcommandx{\customTODO}[3]{\textcolor{#2}{\textsl{#1: #3}}}
\fi
\ifnum\commentType=2
    \usepackage[
    type=upperleft,
    unit=in
    ]{fgruler}
    \usepackage{pdfcomment}
    \newcommandx{\customComment}[3]{\pdfcomment[icon=Comment,opacity=0.5,color=#2,author=#1]{#3}}
    \newcommandx{\customTODO}[3]{\pdfcomment[icon=Note,opacity=0.5,color=#2,author=#1]{#3}}
\fi
\ifnum\commentType=3
    \usepackage[
    type=upperleft,
    unit=in
    ]{fgruler}
    \usepackage{pdfcomment} %
    \usepackage{todonotes} %
    \usepackage{silence}
    \newcommandx{\customComment}[3]{\todo[color=#2!40,size=\small]{\textbf{#1:} #3}}
    \newcommandx{\customTODO}[3]{\todo[color=#2!40,size=\small]{\textbf{#1:} #3}}
\fi

\let\originalleft\left %
\let\originalright\right %
\renewcommand{\left}{\mathopen{}\mathclose\bgroup\originalleft} %
\renewcommand{\right}{\aftergroup\egroup\originalright} %

\definecolor{amber}{rgb}{1.0, 0.49, 0.0}
\definecolor{darkgreen}{rgb}{0.0, 0.5, 0.0}
\definecolor{darkblue}{rgb}{0.0, 0.0, 0.5}
\newcommandx{\All}[1]{\customComment{All}{red}{#1}}
\newcommandx{\Ana}[1]{\customComment{Ana}{amber}{#1}}
\newcommandx{\Justin}[1]{\customComment{Justin}{amber}{#1}}
\newcommandx{\justin}[1]{\customComment{Justin}{amber}{#1}}
\newcommandx{\Vincent}[1]{\customComment{Vincent}{darkblue}{#1}}
\newcommandx{\Oded}[1]{\customComment{Oded}{purple}{#1}}
\newcommandx{\Ahmed}[1]{\customComment{Ahmed}{amber}{#1}}

\newcommandx{\TODO}[1]{\customTODO{TODO}{red}{#1}}
\newcommandx{\AnaTODO}[1]{\customTODO{Ana TODO}{amber}{#1}}
\newcommandx{\JustinTODO}[1]{\customTODO{Justin TODO}{darkgreen}{#1}}
\newcommandx{\OdedTODO}[1]{\customTODO{Oded TODO}{darkgreen}{#1}}

\newcommand{\REMOVE}[1]{} %
\newcommand{\REPLACE}[2]{#2} %

\def\equationautorefname~#1\null{%
  Equation~(#1)\null
}

\newcommand{\eqtag}[1]{{\text{{\small{(#1)}}}}}

\newcommand{\suchThat}[0]{{\mathrm{s.t.}\quad}}

\newcommand{\pos}{\boldsymbol{x}}

\newcommand{\bdr}{\boldsymbol{p}}

\newcommand{\DomainSample}[0]{\pos}
\newcommand{\BoundarySample}[0]{\bdr}

\newcommand{\normal}{{\boldsymbol{n}}}

\newcommand{\Handle}[0]{\boldsymbol{h}}

\newcommand{\bary}[0]{\alpha}
\newcommand{\baryv}[0]{\boldsymbol{\alpha}}

\newcommand{\Net}[0]{\boldsymbol{f}}
\newcommand{\feat}[0]{\boldsymbol{f}}
\newcommand{\vect}[1]{\boldsymbol{#1}}

\newcommand{\fline}[0]{\boldsymbol{\gamma}}

\newcommand{\tpose}[0]{\top}

\newcommand{\Domain}{\mathcal{P}}

\newcommand{\Boundary}{\partial \mathcal{P}}

\newcommand{\R}[0]{{\mathbb{R}}}

\DeclareMathAlphabet{\mathmybb}{U}{bbold}{m}{n}

\newcommand{\Diff}[1]{\,\mathrm{d}#1}

\newcommand{\Laplacian}[0]{\Delta}

\DeclareMathOperator*{\gwn}{wn}

\newcommand{\Hand}{\textsc{Hand}}

\newcommand{\Banana}{\textsc{Banana}}
\newcommand{\Crocodile}{\textsc{Crocodile}}
\newcommand{\Gear}{\textsc{Gear}}
\newcommand{\Shiba}{\textsc{Shiba Inu}}
\newcommand{\Scorpion}{\textsc{Scorpion}}
\newcommand{\ScorpionRand}{\textsc{Scorpion (Randomized)}}
\newcommand{\Beaver}{\textsc{Beaver}}
\newcommand{\Piggybank}{\textsc{Piggybank}}
\newcommand{\Penguin}{\textsc{Penguin}}

\newcommand{\Fish}{\textsc{Fish}}
\newcommand{\Mushroom}{\textsc{Mushroom}}
\newcommand{\Bunny}{\textsc{Bunny}}

\newcommand{\NarrowArrow}{\! \! \rightarrow{} \!}

\gdef\useCroppedImages{1}

\usepackage[export]{adjustbox}
\usepackage{calc}
\usepackage{tabularx}
\usepackage{tikz}
\usepackage{xstring}

\newlength{\beautyHeight}
\newlength{\beautyPixWidth}
\newlength{\beautyPixHeight}
\newlength{\insetvsep}
\gdef\useInsetA{0}
\gdef\useInsetB{0}
\gdef\useInsetC{0}

\newcommand{\setInset}[6]{%
    \expandafter\gdef\csname useInset#1\endcsname{1}%
    \expandafter\gdef\csname inset#1Color\endcsname{#2}%
    \expandafter\gdef\csname crop#1X\endcsname{#3}%
    \expandafter\gdef\csname crop#1Y\endcsname{#4}%
    \expandafter\gdef\csname crop#1W\endcsname{#5}%
    \expandafter\gdef\csname crop#1H\endcsname{#6}%
}

\newcommand{\unsetInset}[1]{%
    \expandafter\gdef\csname useInset#1\endcsname{0}%
}

\newcommand{\addBeautyCrop}[8]{%
    \pdfpxdimen=\dimexpr 1 in/72\relax
    \def\beauty{%
        \let\cropR\relax%
        \let\cropB\relax%
        \newlength\cropR%
        \newlength\cropB%
        \setlength\cropR{{#3 px}-{#5 px}-{#7 px}}%
        \setlength\cropB{{#4 px}-{#6 px}-{#8 px}}%
        \sbox0{\includegraphics[width=#2\textwidth,trim={#5px {\cropB} {\cropR} #6px},clip]{#1}}%
        \begin{tikzpicture}
            \node[anchor=north west,inner sep=0] at (0,0) {\usebox0};
            \begin{scope}[x=\wd0/#7, y=\ht0/#8]
            \if\useInsetA1{
                \draw[\insetAColor,very thick] (\cropAX-#5,-\cropAY+#6) rectangle + (\cropAW,-\cropAH);
            }\fi
            \if\useInsetB1{
                \draw[\insetBColor,very thick] (\cropBX-#5,-\cropBY+#6) rectangle + (\cropBW,-\cropBH);
            }\fi
            \if\useInsetC1{
                \draw[\insetCColor,very thick] (\cropCX-#5,-\cropCY+#6) rectangle + (\cropCW,-\cropCH);
            }\fi
            \end{scope}
        \end{tikzpicture}
    }%
    \setlength\beautyHeight{\heightof{\beauty}}%
    \setlength\beautyPixWidth{#3px}%
    \setlength\beautyPixHeight{#4px}%
    \global\beautyHeight=\beautyHeight%
    \global\beautyPixWidth=\beautyPixWidth%
    \global\beautyPixHeight=\beautyPixHeight%
    \begin{adjustbox}{valign=t}
        \beauty{}
    \end{adjustbox}
}

\newcommand{\trimInset}[6]{%
    \let\cropR\relax%
    \let\cropB\relax%
    \newlength\cropR%
    \newlength\cropB%
    \setlength\cropR{{\beautyPixWidth}-{#3 px}-{#5 px}}%
    \setlength\cropB{{\beautyPixHeight}-{#4 px}-{#6 px}}%
    \color{#2}%
    \fbox{\includegraphics[width=1\linewidth,trim={{#3 px} {\cropB} {\cropR} {#4 px}},clip]{#1}}%
}

\newcommand{\addInset}[2]{%
    \color{#2}%
    \fbox{\includegraphics[width=1\linewidth]{#1}}%
}

\newcommand{\auxtimes}{x}
\newcommand{\auxplus}{+}
\newcommand{\auxspace}{ }

\newcommand{\addInsets}[1]{%
    \begin{adjustbox}{valign=t}
        \StrSubstitute{#1}{.}{-}[\baseFileName]
        \begin{adjustbox}{totalheight=1\beautyHeight,tabular={c}}
            \if\useInsetA1%
                \def\cropfile{\baseFileName-\cropAW\auxtimes\cropAH\auxplus\cropAX\auxplus\cropAY-crop}
                \if\cropInsets1
                    \immediate\write18{convert #1 -crop \cropAW\auxtimes\cropAH\auxplus\cropAX\auxplus\cropAY\auxspace -filter point -resize 800\% \cropfile.png}
                \fi
                \if\useCroppedImages1
                    \addInset{\cropfile.png}{\insetAColor}
                \else
                    \trimInset{#1}{\insetAColor}{\cropAX}{\cropAY}{\cropAW}{\cropAH}%
                \fi%
            \fi%
            \if\useInsetB1%
                \if\useInsetA1\\[\insetvsep]\fi%
                \def\cropfile{\baseFileName-\cropBW\auxtimes\cropBH\auxplus\cropBX\auxplus\cropBY-crop}
                \if\cropInsets1
                    \immediate\write18{convert #1 -crop \cropBW\auxtimes\cropBH\auxplus\cropBX\auxplus\cropBY\auxspace -filter point -resize 800\% \cropfile.png}
                \fi
                \if\useCroppedImages1
                    \addInset{\cropfile.png}{\insetBColor}
                \else
                    \trimInset{#1}{\insetBColor}{\cropBX}{\cropBY}{\cropBW}{\cropBH}%
                \fi%
            \fi%
            \if\useInsetC1%
                \if\useInsetB1\\[\insetvsep]\fi%
                \def\cropfile{\baseFileName-\cropCW\auxtimes\cropCH\auxplus\cropCX\auxplus\cropCY-crop}
                \if\cropInsets1
                    \immediate\write18{convert #1 -crop \cropCW\auxtimes\cropCH\auxplus\cropCX\auxplus\cropCY\auxspace -filter point -resize 800\% \cropfile.png}
                \fi
                \if\useCroppedImages1
                    \addInset{\cropfile.png}{\insetCColor}
                \else
                    \trimInset{#1}{\insetCColor}{\cropCX}{\cropCY}{\cropCW}{\cropCH}%
                \fi%
            \fi%
        \end{adjustbox}
    \end{adjustbox}
}

\definecolor{cartoPrismTeal}{rgb}{0.21960784 0.65098039 0.64705882}
\definecolor{cartoPrismOrange}{rgb}{0.88235294 0.48627451 0.01960784}
\definecolor{cartoPrismGreen}{rgb}{0.45098039 0.68627451 0.28235294}
\definecolor{cartoPrismRed}{rgb}{0.8 0.31372549 0.24313725}
\definecolor{cartoPrismPurple}{rgb}{0.58039216 0.20392157 0.43137255}

\definecolor{mathematicaBlue}{rgb}{0.38, 0.51, 0.71}
\definecolor{mathematicaOrange}{rgb}{0.88, 0.61, 0.14}
\definecolor{mathematicaGreen}{rgb}{0.56, 0.69, 0.19}
\definecolor{mathematicaRed}{rgb}{0.92,0.39, 0.21}
\definecolor{mathematicaPurple}{rgb}{0.53, 0.47, 0.7}

\newtheorem*{remark}{Remark}

\newcommand{\anonimg}[2]{%
    \makeatletter%
    \if@ACM@anonymous%
        \includegraphics[width=\linewidth]{#1}%
    \else%
        \includegraphics[width=\linewidth]{#2}%
    \fi%
    \makeatother%
}
\usepackage{enumitem,enumerate}
\setlist[itemize]{noitemsep, nolistsep, leftmargin=*}

\setcopyright{cc}
\setcctype{by-nc-sa}
\acmJournal{TOG}
\acmYear{2025} \acmVolume{1} \acmNumber{1} \acmArticle{1} \acmMonth{1} \acmPrice{}\acmDOI{10.1145/3771928}

\begin{document}

\title[Robust Biharmonic Skinning Using Geometric Fields]{Robust Biharmonic Skinning Using Geometric Fields}

\author{Ana Dodik}
\email{anadodik@mit.edu}
\orcid{1234-5678-9012}
\affiliation{%
  \institution{MIT CSAIL}
  \country{USA}
  \city{Cambridge}
}

\author{Vincent Sitzmann}
\orcid{1234-5678-9012}
\affiliation{%
  \institution{MIT CSAIL}
  \country{USA}
  \city{Cambridge}
}

\author{Justin Solomon}
\orcid{1234-5678-9012}
\affiliation{%
  \institution{MIT CSAIL}
  \country{USA}
  \city{Cambridge}
}

\author{Oded Stein}
\orcid{1234-5678-9012}
\affiliation{%
  \institution{University of Southern California and MIT CSAIL}
  \country{USA}
  \city{Los Angeles}
}

\begin{abstract}
Bounded bihramonic weights are a popular tool used to rig and deform characters for animation, to compute reduced-order simulations, and to define feature descriptors for geometry processing.
They necessitate tetrahedralizing the volume bounded by the surface, introducing the possibility of meshing artifacts or  tetrahedralization failure.
We introduce a \emph{mesh-free} and \emph{robust} automatic skinning technique that generates weights comparable to the current state of the art, but works reliably even on open surfaces, triangle soups, and point clouds where current methods fail.
We achieve this through the use of a specialized Lagrangian representation enabled by the advent of hardware ray-tracing, which circumvents the need for finite elements while optimizing the biharmonic energy and enforcing boundary conditions.
The flexibility of our formulation allows us to integrate artistic control through weight painting during the optimization.
We offer a thorough qualitative and quantitative evaluation of our method.
\end{abstract}
\begin{CCSXML}
<ccs2012>
<concept>
<concept_id>10010147.10010371.10010396.10010398</concept_id>
<concept_desc>Computing methodologies~Mesh geometry models</concept_desc>
<concept_significance>500</concept_significance>
</concept>
<concept>
<concept_id>10010147.10010371.10010396.10010397</concept_id>
<concept_desc>Computing methodologies~Mesh models</concept_desc>
<concept_significance>500</concept_significance>
</concept>
</ccs2012>
<concept>
<concept_id>10010147.10010371.10010352</concept_id>
<concept_desc>Computing methodologies~Animation</concept_desc>
<concept_significance>500</concept_significance>
</concept>
<concept>
<concept_id>10010147.10010371.10010396.10010400</concept_id>
<concept_desc>Computing methodologies~Point-based models</concept_desc>
<concept_significance>500</concept_significance>
</concept>
<concept>
<concept_id>10010147.10010341</concept_id>
<concept_desc>Computing methodologies~Modeling and simulation</concept_desc>
<concept_significance>500</concept_significance>
</concept>
<concept>
<concept_id>10010147.10010371.10010372.10010374</concept_id>
<concept_desc>Computing methodologies~Ray tracing</concept_desc>
<concept_significance>300</concept_significance>
</concept>
<concept>
<concept_id>10010147.10010371.10010372.10010377</concept_id>
<concept_desc>Computing methodologies~Visibility</concept_desc>
<concept_significance>300</concept_significance>
</concept>
\end{CCSXML}

\ccsdesc[500]{Computing methodologies~Mesh geometry models}
\ccsdesc[500]{Computing methodologies~Mesh models}
\ccsdesc[500]{Computing methodologies~Animation}
\ccsdesc[500]{Computing methodologies~Point-based models}
\ccsdesc[500]{Computing methodologies~Modeling and simulation}
\ccsdesc[300]{Computing methodologies~Ray tracing}
\ccsdesc[300]{Computing methodologies~Visibility}
\keywords{skinning weights, bounded biharmonic weights, geometry processing, deformation, partial differential equations, variational problems}

\begin{teaserfigure}
  \centering  
  \includegraphics[width=\linewidth]{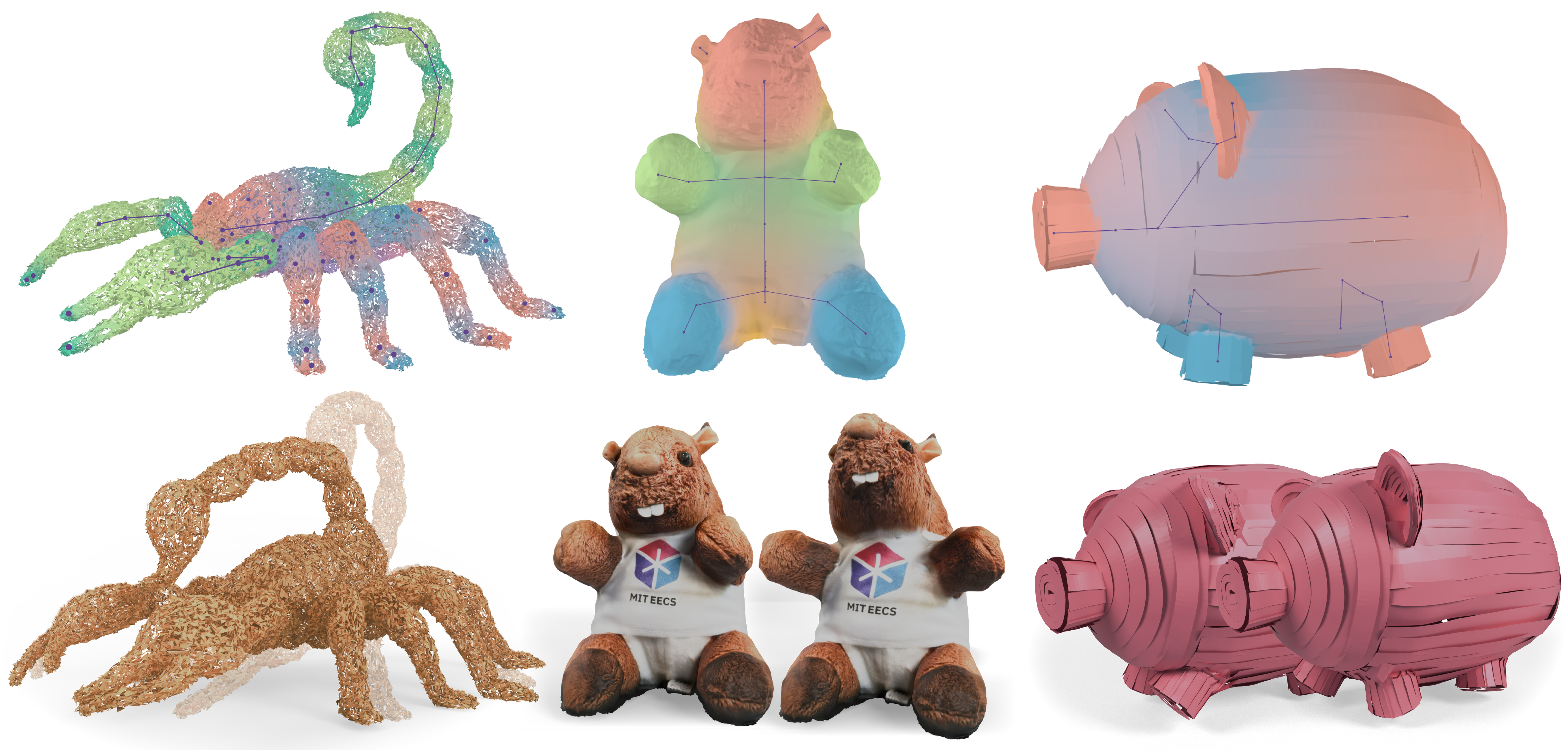}
  \caption{Bounded biharmonic weights produced by our method can be used to extrapolate color (top row) or deformations (bottom row) from control handles. Our method succeeds on meshes that are typically difficult to handle with standard tools. As an extreme example, the \ScorpionRand{} triangle soup is missing $30$\% of its faces and has had the remaining ones randomly perturbed. We can robustly compute weights for meshes created using off-the-shelf $3$D scanning software even if they are non-watertight and contain thin and self-intersecting triangles as does the \Beaver{} mesh. 
  Similarly, our method can be used to deform virtual reality ribbon drawings \cite{rosales2019SurfaceBrush}, such as the \Piggybank{} mesh.}
  \Description{Three different meshes are shown, top row contains the meshes with skinning weights visualized in various colors, bottom row shows animations of the meshes in the top row.}\label{fig:teaser}
\end{teaserfigure}  

\maketitle

\section{Introduction}

The dominant pipeline for computer animation relies on \emph{deformation skeletons} composed of control handles (typically points and bones), each of which has an associated region of influence on the shape.
These regions of influence, referred to as \emph{skinning weights}, specify how deformations of the control handles are to be blended and transferred to the shape.
The skeleton interface requires users to specify the skinning weights, traditionally through the slow and tedious process of manualweight painting .

To ease this burden, various automatic skinning weight computation methods have been developed.
Despite growing interest, data driven approaches remain class-specific, require an abundance of data, and struggle with out-of-distribution examples.
Instead, a popular alternative is to formulate skinning weights as minimizers of some smoothness objective, side-stepping generalization issues of data-driven methods (\S\ref{ss:prevskinning}).
Such optimization-based approaches have seen adoption in commercial software, notably, Pinnochio \cite{baran2007automatic} in Blender and bounded biharmonic weights (BBW) \cite{BBW:2011} in Adobe Character Animator.

\begin{figure}[ht!]
  \centering
  \includegraphics[width=\linewidth]{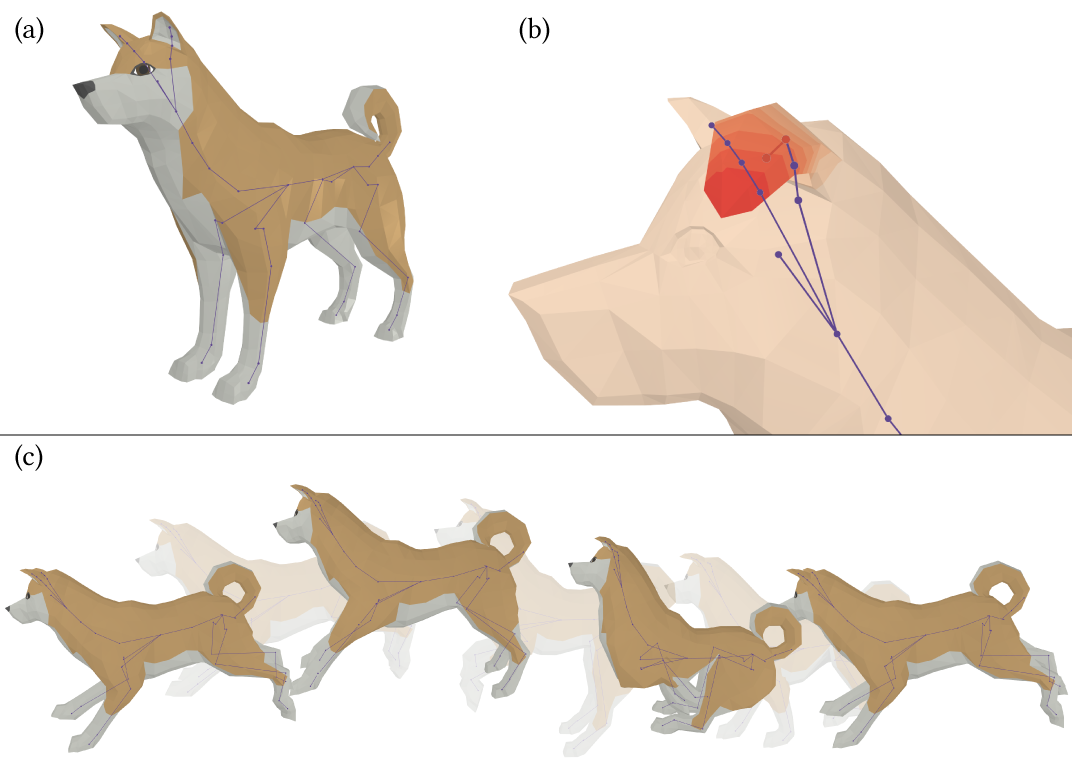}
  \caption{An illustrative example on the low-poly \Shiba{} mesh, with the skeleton depicted in purple (a). The skinning weight function of the ear bone determines the region of influence of that bone (b). Our method for skinning weight computation produces smooth-looking deformations (c).}
  \Description{A mesh of a dog, visualized skinning weights, and animation of the mesh.}\label{fig:shiba}
\end{figure}

Despite this adoption, manual painting of skinning weights remains predominant.
This is, in part, due to \textbf{robustness} issues with automatic methods, as they rely on mesh-based discretizations using the finite-element method (FEM).
Famously, FEM-based approaches struggle with data common in the real world \cite{sharp2020flipout, sharp2020laplacian, Sharp:2021:GPI, gillespie2021integer, triwild, Sawhney:2020:MCG, Sawhney:2022:DND, Miller:2023:BVC, Sawhney:2023:WoSt}.
For example, Blender's implementation raises the error \texttt{Bone Heat Weighting: Failed to find solution...} due to issues with FEM, resulting in a plethora of online discussions and tutorials dedicated to cleaning up meshes to avoid this error.

Robustness issues are especially pronounced with methods that rely on tetrahedral meshes, as fast and robust tetrahedralization remains a challenging open problem \cite{tetwild, ftetwild, Diazzi:2023}.
This is particularly unfortunate, as the state-of-the-art smoothness-based  formulation of skinning weights---bounded biharmonic weights \cite{BBW:2011}---requires a tetrahedral mesh.
Figure~\ref{fig:tetfail} demonstrates a common failure case of state-of-the-art fast tetrahedralization software~\cite{ftetwild}.
If we want to guarantee that our software can compute a tetrahedral mesh, we either get a fast approximate solution that can lead to visible artifacts, or the tetrahedralization software can take multiple hours only for the downstream FEM solver to fail due to the tetrahedral mesh being too high-resolution.

In addition, manual weight painting remains prevalent because it offers complete \textbf{artistic control} over the final look of the animation.
Therefore, instead of trying to fully automate the entire skinning pipeline, automatic skinning methods need to offer ways to integrate artistic edits.

Recent function representations popularized in machine learning and vision promise to replace FEM in optimization pipelines for  geometry processing tasks while introducing additional artistic control \cite{Dodik:VBC:2023}.
However, representations like neural fields \cite{xie2022neural} or Gaussian splatting \cite{Kerbl2023} are not directly applicable to geometry processing due to hard constraints on the set of permissible solutions as dictated by the geometry of the shape (Fig.~\ref{fig:rt}, \S\ref{ss:kernel}), boundary conditions (\S\ref{ss:bc}) or the problem formulation itself (Eq.~\ref{eq:bbw}).
We take inspiration from Gaussian splatting but incorporate hard constraints directly into the architecture of the differentiable field, dubbing our construction \emph{geometric fields}.

In this article, we introduce a mesh-free, tailor-made representation for skinning weights that satisfies the constraints of the problem by construction.
Our representation not only accelerates weights computation, but also respects the constraints imposed by the geometry of the shape, avoiding artifacts like bleeding (Fig.~\ref{fig:rt}). 
With our architecture, we can optimize for bounded biharmonic weights  via stochastic gradient descent (SGD).

Our approach inherits other benefits of differentiable programming and gradient-based optimization.
For example, we can incorporate user-painted weights, offering additional control over the output. %
Instead of requiring the user to paint the entire shape, our method enables weight painting on a subset of the boundary, which are then included as Dirichlet boundary conditions in the optimization problem.
Our system quickly produces an initial solution that a user can paint over according to preference---and then resume the optimization to incorporate the prescribed weights into the solution.

Our weights are universally faster to compute than the original BBW implementation \cite{BBW:2011}, while fast quasi-harmonic weights (QHW) \cite{wang2021fast} are faster only on well-behaved meshes.
As soon as robust tetrahedralization is needed, our method is more robust and faster end-to-end than existing alternatives. In many cases, tetrahedralization can entirely fail or produce unusable results (Figure~\ref{fig:tetfail}); in cases where it succeeds, QHW can crash or be slower than our method (Table~\ref{tbl:qant}, Figure~\ref{fig:timings}). 

In summary, our main contributions are:
\begin{itemize}
  \item A robust automatic method for computing skinning weights.%
  \item A Lagrangian representation and approach to solving the bounded biharmonic %
  weights problem without finite elements.
  \item The use of hardware-accelerated ray tracing for a geometry-aware function parameterization. %
  \item Modifications that allow the method to work on non-watertight domains like triangle soups, ribbon drawings, or point clouds. %
  \item A way of incorporating weight painting into the optimization by introducing Dirichlet boundary conditions.
  \item A thorough qualitative and quantitative investigation of our method.
\end{itemize}

\section{Related Work}\label{ss:relatedwork}

In this section, we mention connections of our work to other methods in the geometry processing literature.

\subsection{Robust Geometry Processing}

A significant body of prior work has been dedicated to making FEM-based geometry processing robust.
FEM-based algorithms put the burden on the end-user to ensure that meshes are ``well-behaved,'' requiring the user to e.g.\ remove self-intersections or to ensure mathematical properties such as manifoldness, watertightness, or boundedness of interior angles.
While FEM can be a natural fit for meshes, a mesh that is sufficient for representing a shape is not necessarily also a good FEM mesh; moreover, FEM-based methods often require meshing the interior of a volume bounded by a boundary representation, while the latter is sufficient for 3D modeling.

\begin{figure}[t!]
  \centering
  \includegraphics[width=\linewidth]{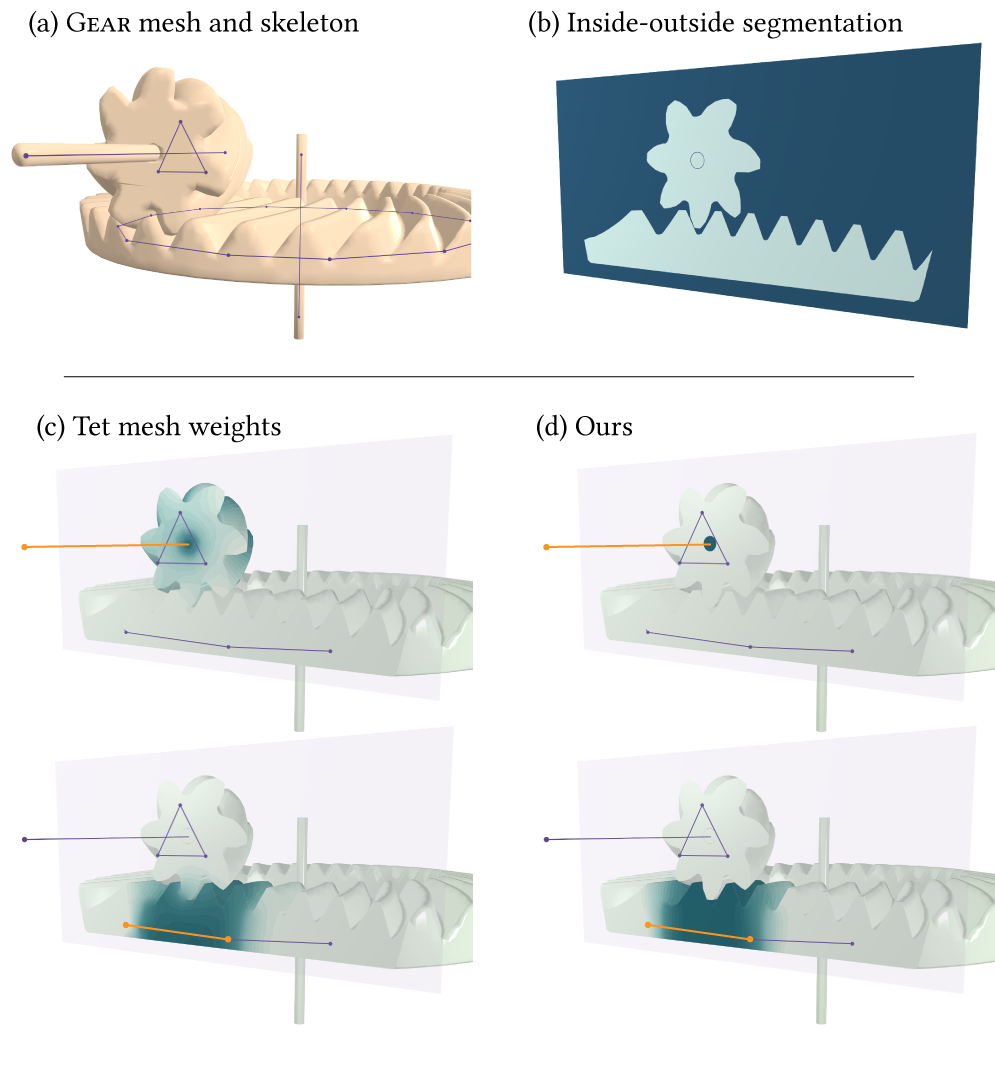}
  \caption{Our method for bounded biharmonic weights \cite{BBW:2011} works on poorly behaved geometry such as the challenging \Gear{} mesh (a). 
  We visualize quantities as colors on a planar slice through the volume. 
  The mesh does not self-intersect, as evidenced by the generalized winding number \cite{Jacobson2013Winding} (b).
  Solving for BBW using on a tetrahedral mesh from \textsc{FastTetWild} \cite{ftetwild} with default parameters results in weights bleeding over boundaries and discretization artifacts (c).
  Modifying \textsc{FastTetWild}'s parameters to respect boundaries increases computation time to $\mathbf{1.78}$ \textbf{hours}.
  In comparison, our weights are smooth, respect boundaries, and can be computed without tetrahedralization in \textbf{$\mathbf{32.2}$ seconds} (d).}
  \Description{A mesh with many fine filigree-like details.}\label{fig:tetfail}
\end{figure}

Robust triangle meshing algorithms rely on complex data structures and often parallelize poorly \cite{sharp2020flipout, sharp2020laplacian, Sharp:2021:GPI, gillespie2021integer, triwild}.
Existing tetrahedral meshing algorithms either have strong requirements on the quality of the input (e.g., crashing in the presence of self-intersections) \cite{cgal:tet, Diazzi:2023} or drastically change the appearance of the mesh \cite{tetwild, ftetwild} as in Figure~\ref{fig:tetfail}.

To sidestep the issues above, there is emerging interest in mesh-free approaches to geometry processing using Monte Carlo methods \cite{Sawhney:2020:MCG, Sawhney:2022:DND, Miller:2023:BVC, Sawhney:2023:WoSt}.
As of now, this approach is not applicable to our setting, as it applies to linear PDEs rather than the inequality-constrained variational problem needed for skinning.
Similarly, the boundary element method (BEM) holds some promise for alleviating dependence on tetrahedralization of a boundary representation, but typical BEM algorithms are restricted to linear PDE in the interior of the domain.

A seminal construction in robust geometry processing, the generalized winding number gives a robust means of deciding whether a point is inside or outside a given shape~\cite{Jacobson2013Winding}.
It degrades gracefully and produces meaningful results even in the presence of heavily degenerate boundary geometry.
Its main application has been to approximate tetrahedral meshing methods, such as the one used to create Figure~\ref{fig:tetfail}. We use the generalized winding number to alleviate the need for tetrahedralization while evaluating our optimization objective (see Section~\ref{ss:gwn}).

\subsection{Automatic Skinning Weights}\label{ss:prevskinning}

In an effort to alleviate the largely manual process of painting skinning weights, automatic skinning weights algorithms use input geometry and/or example poses to infer the weight functions.  This paper focuses on the former setting---by far the most common, since obtaining an exemplary set of poses for a shape that suggest the underlying skinning weights is itself a hard problem.  See e.g.\ \cite{james2005skinning,kavan2010fast,le2012smooth,le2014robust,wampler2016fast} for data-driven methods in the latter category; we also mention some neural network architectures that predict skinning weights below.

\paragraph*{PDE-based methods.} Many classical methods for automatic skinning weights are built around the solution of a partial differential equation (PDE) such as the Laplace equation $\Delta w\equiv0$ \citep{joshi2007harmonic}, which can be efficiently solved with Dirichlet boundary conditions and whose solutions automatically fulfill the constraints needed for skinning.
The related \emph{Pinocchio} method \cite{baran2007automatic} is implemented in common modeling software like Blender and Maya. %
Low-order PDEs like the Laplace equation can lead to artifacts near control handles.
This is alleviated by solving the higher-order biharmonic equation $\Delta^2\alpha\equiv0$ \citep{botsch2004intuitive} or even the triharmonic equation \citep{tosun2008geometric,jacobson2010mixed}.
These methods, however, lose the maximum principle that holds for the Laplace equation, leading to undesirable oscillatory skinning weights.%

\paragraph*{Variational methods.}
Variational methods add constraints on top of a PDE-derived energy, making the problem nonlinear.
The widest-known and state-of-the-art method of this class is \emph{bounded biharmonic weights} (BBW) \citep{BBW:2011}, which uses a biharmonic energy with nonnegativity and partition of unity constraints---and an implicit Neumann boundary condition~\citep{stein2018natural}---leading to a convex quadratic program that can be discretized on a triangle/tetrahedral mesh.
BBW provides high-quality, smooth automatic skinning weights at the cost of efficiency and mesh dependence.
It has been extended to avoid local extrema \citep{jacobson2012smooth} or to improve efficiency \citep{wang2021fast}.

These methods, however, still rely on meshing the entire computational domain (a volume in 3D), even though skinning weights are only needed at the outer surface;
it is unclear how to extend them to domains that cannot easily be meshed, e.g., because they are not watertight. 
\citet{Jacobson2013Winding} briefly illustrate how generalized winding numbers---used as a means of distinguishing inside from outside in non-watertight models---can possibly bridge the gap (see their Figure 20).  Our extension to non-watertight domains in Section~\ref{ss:gwn} also makes use of \citet{Jacobson2013Winding}'s model.

\paragraph*{Additional approaches.}  A zoo of methods propose models adjacent to automatic skinning weights computation. For example, \citet{thiery2018araplbs} jointly optimize bone positions and skinning weights using models from elasticity.  Delta mush methods~\cite{le2019direct} bypass steps of skinning and weight computation by applying \emph{a posteriori} smoothing to linear-blend skinning deformations using binary binding weights. \citet{bang2018spline} propose a spline-based interface to efficiently create and edit skinning weights.

\citet{dionne2013geodesic,dionne2014geodesic} consider the problem of computing robust skinning weights using a closed-form formula in terms of distances; they use a variant of Dijkstra's algorithm to compute distances on a voxel grid of the interior of the domain.  Their method is robust to non-watertight edges but requires a dense voxelixation in the presence of thin features to avoid bleeding between disconnected parts of the shape; they also do not optimize a smoothness energy, which can yield artifacts at points where shortest-path distance is not differentiable. 
 \citet{xian2018efficient} use a similar approach for 2D image deformation but compute interior distances along a graph of mesh edges.

\paragraph*{Neural representations.}  \citet{Dodik:VBC:2023}'s method to compute generalized barycentric coordinates using a neural representation is closely related to our approach; see their work for for past literature on generalized barycentric coordinates.  Although the idea of using machine learning-inspired function representations for geometry processing also underlies our work, their method is not directly applicable to our problem. In particular, building in hard constraints of the function class of barycentric coordinates requires a complicated purpose-built architecture that is unnecessary in our formulation  (see Section~\ref{ss:kernelparam}).

Other works use neural networks to predict skinning weights learned from data.
These methods are \emph{not} comparable to ours, since they are data-driven and typically omit the necessary hard constraints, although their architectures parameterize skinning weights functions in various ways, providing a point of high-level comparison.
Several axes can be used to compare methods for learning skinning weights.  A key decision is the architecture and its interaction with the shape representation; for example, many learned skinning weights employ graph neural networks \cite{liu2019neuroskinning, pan2021heterskinnet,mosella2022skinningnet}, while others use convolutional neural networks \cite{ouyang2020autoskin} or Gaussian splatting \cite{kocabas2023hugs} to represent and process shapes.  Another axis is the training data. While many works rely on direct supervision, some infer skinning weights from deformations or by jointly optimizing bone positions with skinning weights \cite{yang2021s3,ma2023tarig,chen2021snarf,li2021learning,xu2020rignet}; others incorporate modalities like video \cite{liao2023vinecs}. \citet{jeruzalski2020nilbs,kant2023invertible} use neural representations to solve inverse skinning problems.  A few methods also introduced specialized architectures and loss terms, e.g., for handling interactions between cloth and human bodies \cite{wu2020skinning,ma2022neural} or for supporting a pipeline to learn articulated 3D animals \cite{wu2023magicpony}.  Beyond using Laplacian regularization  (see e.g.\ \cite{liao2023vinecs}), these papers primarily infer skinning weights from data rather than using neural representations to optimize for geometric skinning weights.

\subsection{Physics Informed Neural Networks in Graphics}

A recently popular approach for discretization-free PDE solutions on geometric domains employs physics-informed neural networks (PINNs), such as the works of \citet{raissi2019physics}.
These are used, e.g., to solve PDEs for physical simulation \cite{Chen2023INSR}, find watertight surfaces from point clouds \cite{gropp2020implicit,sitzmann2019siren}, and to solve general hyperbolic equations \cite{rodrigueztorrado2021physicsinformed}.
Similar to the way we enforce boundary conditions, these networks can be enhanced with explicit boundary constraints \cite{lu2021hard,Sukumar2022,Chen2024ahard,anonymous2024scaling}.
Other approaches for enforcing constraints on PDEs include \citep{djeumou2022neural,zhong2023neural,Liu2022aunified,Mohan2023}.

\section{Robust Biharmonic Skinning}\label{s:prelim}

Suppose we are given a shape $\Domain \subset \R^d$ for $d$ either $2$ or $3$, \linebreak with boundary $\partial \Domain$ and a skeleton consisting of $K$ \emph{control handles} ${\left\{ \Handle_i \subset \Domain \colon 1 \leq i \leq K \right\}}$.
Most commonly, control handles are either points ($\Handle_i$ is a singleton set) or bones ($\Handle_i$ is a line-segment).
Each  handle has an associated \emph{skinning weight function}, $\bary_i : \Domain \rightarrow [0, 1]$, with ${\baryv(\DomainSample) = {\left[\bary_1(\DomainSample), \ldots, \bary_K(\DomainSample)\right]}^\tpose}$ denoting the vector-valued skinning weights function.

Skinning weights functions %
satisfy a set of properties at each point $\DomainSample \in \Domain$:
\begin{itemize}
  \item \textsc{Non-negativity.} A handle cannot have negative influence over a point: $\bary_i(\DomainSample) \geq 0$.
  \item \textsc{Partition of unity.} The influence of all handles over any point must sum up to $100\%$: $\sum_i\bary_i(\DomainSample) = 1$. 
  \item \textsc{Lagrange property.} A handle must have $100\%$ influence over itself: $\forall \DomainSample \in \Handle_j \colon \bary_i(\DomainSample) = \delta_{ij}$, where $\delta_{ij}$ is the
  Kronecker delta.%
  \item \textsc{Boundary conditions.} Optionally, a model for skinning weights can include boundary conditions that specify the behavior of the weights at $\partial\Domain$, e.g., prescribing the value of the normal derivative at the boundary $\normal_{\bdr}^\tpose \nabla \bary_i(\BoundarySample) = 0$ via Neumann conditions. We introduce the ability to optionally prescribe the value of the weights on a subset of the boundary, $\bary_i|_{\Gamma \subseteq \partial\Domain}$ via Dirichlet conditions. %
\end{itemize}
Many functions $\baryv$ satisfy these constraints, so algorithms for skinning weights computation typically seek weights that extremize a functional like smoothness.

\begin{figure}[t!]
  \centering
  \includegraphics[width=\linewidth]{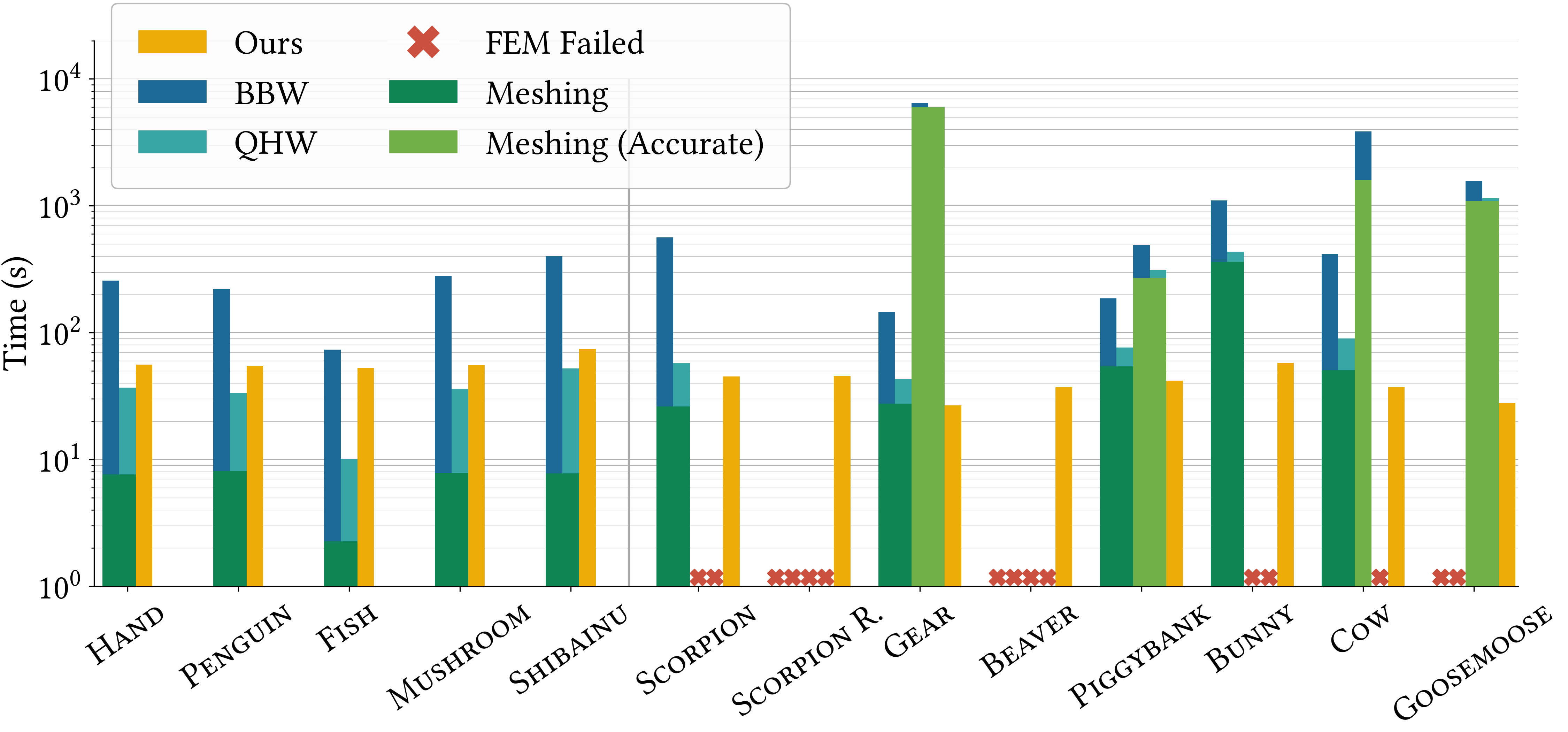}
  \caption{On well behaved meshes, our method is faster than BBW and somewhat slower than QHW (left). On meshes requiring robust tetrahedralization, our method succeeds in all test cases, whereas existing alternatives either fail to produce a solution or are up to orders of magnitude slower (right).}
  \Description{A bar plot showing the timings for different meshes, comparing our method with BBW, QHW using different tetrahedral meshing techniques.}\label{fig:timings}
\end{figure}

Although state-of-the-art algorithms that optimize \eqref{eq:bbw} are fairly efficient, they necessitate tetrahedral meshing, which forces a compromise between boundary approximation quality (Figure~\ref{fig:tetfail}), runtime (Figure~\ref{fig:timings}, Table~\ref{tbl:qant}), and/or manual intervention to repair bad boundary elements. %
Once the tetrahedral mesh is generated, success is still not guaranteed thanks to bad mesh elements and/or poorly-scaling optimization algorithms (Figure~\ref{fig:timings}, red crosses). %

To address this challenge, the key idea of this work is to define the space of \emph{geometric fields}---smooth parametric functions which incorporate into their definition the geometry of a given shape, as well as necessary hard constraints and boundary conditions.
Higher-order differentiability allows us to optimize for smoothness using stochastic gradient descent and to easily incorporate artistic edits into the optimization (Figures \ref{fig:goosemoose}, \ref{fig:cow}).

To make geometric fields conform to the interior of a shape, we design a geometry-aware point-based formulation that exploits modern ray-tracing hardware (\S\ref{ss:kernelparam}), making them robust to boundaries with bad or non-manifold mesh elements and self-intersections.
As a consequence, our method inherits the practically sub-linear scaling of ray-tracing with respect to boundary resolution.

We showcase geometric fields by robustly solving  the seminal \emph{bounded biharmonic weights} skinning weights problem \cite{BBW:2011}. Combining the constraints and objectives above into a single optimization problem, \citet{BBW:2011} solve:
\begin{subequations}\label{eq:model}
\setcounter{equation}{-1}%
\renewcommand{\theequation}{\theparentequation}%
\begin{flalign}
    && \min_{\bary_{i \in [1, K]}} &
        \sum_{i=1}^{K} \int_\Domain \vert \Laplacian \bary_i(\DomainSample) \vert^2 \Diff V(\DomainSample),& \label{eq:bbw}\\
    && \suchThat & \bary_i(\DomainSample) \geq 0\,,\;\forall i, \DomainSample \in \Domain, & \eqtag{non-negativity} \label{eq:nonneg} \tag{\theequation{}.1} \\[5pt]  
    &&& {\textstyle\sum_i} \bary_i(\DomainSample) = 1\,,\;\forall \DomainSample \in \Domain, & \eqtag{partition of unity} \label{eq:partition} \tag{\theequation{}.2} \\[5pt]
    &&& \bary_i(\DomainSample) = \delta_{ij}\,,\;\forall i, \DomainSample \in \Handle_j, & \eqtag{Lagrange property} \label{eq:lagrange} \tag{\theequation{}.3}
\end{flalign}
\end{subequations}
where $\delta_{ij}$ is the Kronecker delta, $V$ is the volume form of $\Domain$.

\paragraph*{Boundary Conditions} The constraints in Equations~\ref{eq:nonneg},~\ref{eq:partition},~\ref{eq:lagrange} can further be combined with boundary conditions for the BBW variational problem.
In addition to natural boundary conditions, our method can be modified to support zero Neumann conditions or Dirichlet conditions defined on a subset of the boundary as a way of incorporating user-defined brush strokes into the skinning weights computation.
Formally, a user can choose to enable either of the following boundary conditions,
\begin{subequations}\label{eq:modelbc}
\setcounter{equation}{-1}%
\renewcommand{\theequation}{\theparentequation}%
\begin{flalign}
    &&& \normal_{\bdr}^\tpose \nabla \bary_i(\BoundarySample) = 0\,,\;\forall i, \BoundarySample \in \Gamma_1 \subseteq \Boundary, & \eqtag{zero Neumann} \tag{\ref{eq:bbw}.4a} \label{eq:zeroneum} \\[5pt]
    &&& \baryv(\BoundarySample) = \vect{g}(\BoundarySample) \,,\;\forall \BoundarySample \in \Gamma_2 \subseteq \Boundary, & \eqtag{Dirichlet} \tag{\ref{eq:bbw}.4b} \label{eq:dirichlet}
\end{flalign}
\end{subequations}
\addtocounter{equation}{-1}%
where $\Gamma_1$ and $\Gamma_2$ are non-overlapping subsets of the boundary, $\normal_{\bdr}$ is the surface normal at the boundary point $\BoundarySample$, and $\vect{g}$ represents user-drawn Dirichlet conditions. 
Note that, while the original BBW article~\cite{BBW:2011} does not explicitly employ zero Neumann constraints,~\citet{stein2018natural} show that their discretization induces these conditions (see their \S2.2.1). 
The choice of boundary conditions is subjective, and our system is flexible enough to turn 
boundary conditions on or off as desired---even during optimization---allowing users to manually prescribe weights at certain locations.

\section{Geometric Fields}\label{s:gemfields}

We begin by presenting the design of a flexible differentiable field architecture that respects the geometry of the shape and satisfies the hard constraints and boundary conditions necessary for skinning.
At a high level, our model composes a smooth parametric \emph{geometry-aware} function ${\Net_{\theta} : \Domain \rightarrow \R^K}$, which maps every point in the domain to one value per control handle (\S\ref{ss:kernel}), with different \emph{activation functions} that ensure the output satisfies constraints and boundary conditions (\S\ref{ss:activation},\ref{ss:bc}).

\subsection{Function Representation}\label{ss:kernel}

In this section, we formulate a geometry-aware parametric function representation, $\Net_{\theta}$, which we will use to represent skinning weights.
It is tempting to use a neural field as a general-purpose function representation, yet these are typically unaware of a shape's geometry.
Instead, we propose a more interpretable and geometry-aware representation inspired by classical computer graphics models as well as modern splat-based rendering methods.

\begin{figure}[h]
  \centering
  \includegraphics[width=\linewidth]{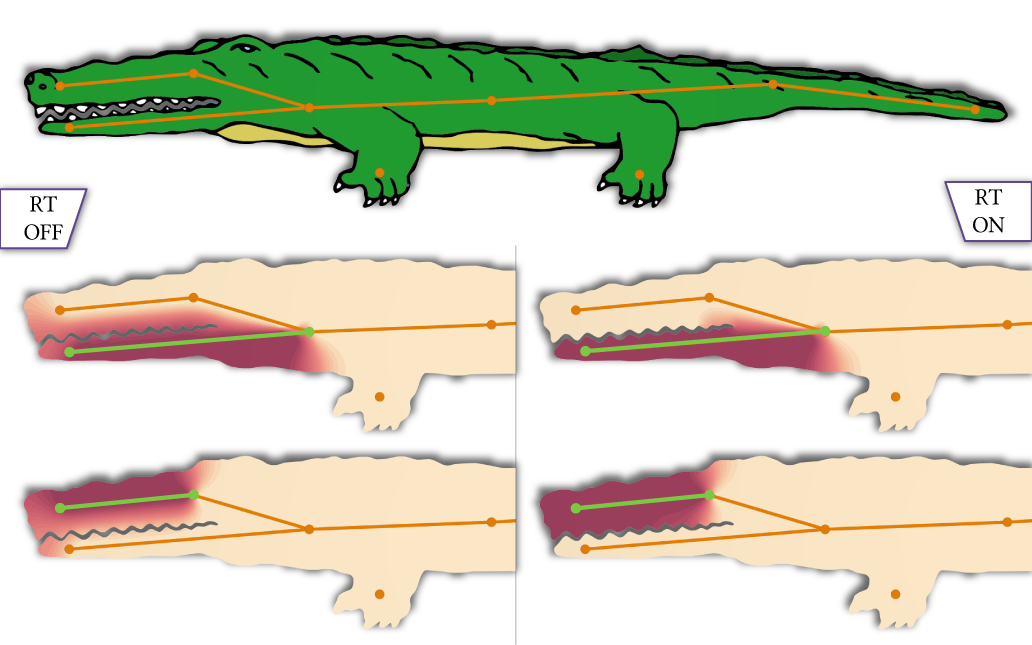}
  \caption{We demonstrate the effects of our visibility-aware kernel on the \Crocodile{} mesh. On the left, skinning weights optimized using $k_\textsc{e}$ exhibit undesirable \emph{bleeding artifacts}. On the right, we can see that optimizing for skinning weights using $k_\textsc{rt}$ mitigates the issue.}
  \Description{A 2D Crocodile mesh with its jaws being close together with one skeleton bone on each side of its jaws.}\label{fig:rt}
\end{figure}

\paragraph*{Kernel-Based Parameterization}\label{ss:kernelparam}

We parameterize $\Net_{\theta}$ using a collection of pairs 
${\mathcal{X} \coloneqq {\left\{ \left( \DomainSample_i, \feat_i \right) \colon 1 \leq i \leq N\right\}}}$, where $\DomainSample_i \in \Domain$ is a randomly sampled point in the domain and $\feat_i \in \R^K$ is a parameter vector associated to that point.
The points $\DomainSample_i$ are fixed during optimization; the optimization variables are the parameter vectors $\feat_i$ , which we can think of as rows of $\theta \in \R^{N \times K}$.
Section~\ref{ss:est} and Appendix~\ref{s:impl} detail how the $\DomainSample_i$'s are sampled.

Evaluated at a point $\DomainSample \in \Domain$, we take %
\begin{equation}~\label{eq:rawinterp}
  \Net_{\theta}(\DomainSample) \coloneqq {\frac{\sum_{i=1}^{N}{k(\boldsymbol{x}, \boldsymbol{x}_i) \boldsymbol{f}_i}}{\sum_{i=1}^{N}{k(\boldsymbol{x}, \boldsymbol{x}_i)}}},
\end{equation}
where $k(\boldsymbol{x}, \boldsymbol{x}_i)$ is an affinity kernel between $\boldsymbol{x}$ and $\boldsymbol{x}_i$.

This point-based form for $\Net_{\theta}$ is built on classical methods for kernel regression \cite{Nadaraya:1964, Watson:1964} and is a common tool in computer graphics \cite{CoifmanLafon:2006, Pauly:2003, Alexa:2004, gingold1977smoothed,lucy1977numerical}. %
Unlike these formulations, which are oblivious to the boundary of $\Domain$, relying on modern machine learning techniques and advances in GPU hardware lets us design a kernel $k$ that transforms this into a viable representation for variational problems in geometry.

Introducing a kernel-based formulation rather than a neural network is justified by several factors.
Bounded biharmonic weights are smooth, meaning they can be well-approximated by samples on a point cloud. 
Furthermore, this form for $\Net_{\theta}$ localizes the degrees of freedom on the domain, making it possible to visualize and reason about design parameters as shown in \S\ref{s:results} and Appendix~\ref{s:impl}.
Most importantly, this formulation admits for a simple modification to the kernel $k$ that incorporates structural information and prevents \emph{bleeding} over the boundaries of the shape, shown in Figure~\ref{fig:rt}.
Avoiding bleeding artifacts would be non-trivial using a coordinate neural network, since a coordinate network in its usual configuration is extrinsic by nature---information that is far away intrinsically can be close together extrinsically (for example, the raised arm of the robot and its head in Figure-\ref{fig:soup}), and an extrinsic coordinate network will cause bleeding across narrow gaps.

\paragraph*{Visibility-based Kernel}~\label{ss:topo} A standard choice for an affinity kernel is the exponential kernel~\cite{Nadaraya:1964, Watson:1964, CoifmanLafon:2006}:
\begin{equation}~\label{eq:kernel}
  k_{\textsc{e}}(\boldsymbol{x}, \boldsymbol{x}_i) = \exp\left\{ -\frac{1}{2 \sigma^2} \left\Vert \boldsymbol{x} - \boldsymbol{x}_i \right\Vert^2 \right\},
\end{equation}
where $\sigma$ governs the \emph{spread} of the kernel.
However, as this kernel is based on the extrinsic distance between $\vect{x}$ and $\vect x_i$, it can lead to bleeding artifacts demonstrated in Figure~\ref{fig:rt}.

Previous work \cite{BBW:2011, Jacobson:MONO:2012, wang2021fast} avoids bleeding by meshing the interior of $\Domain$ and associating degrees of freedom and objective terms with elements of the mesh.
Reliance on a conforming mesh  with sufficient quality to discretize and solve skinning weights problems hamstrings the efficiency and reliability of automatic skinning weights computation, since design of fast and robust meshing algorithms remains challenging \cite{tetwild, ftetwild, Diazzi:2023}.  Moreover, this limitation prevents computation of skinning weights on triangle soups and other  disconnected domains, as we consider in Section~\ref{ss:gwn}.

To address bleeding using GPU-friendly queries without a volume mesh, we modify the kernel $k$ to only allow for pairs of points if they are \emph{visible} to each other:
\begin{equation}~\label{eq:kernelrt}
  k_{\textsc{rt}}(\boldsymbol{x}, \vect x_i) = \mathcal{V}(\boldsymbol{x} \leftrightarrow \vect x_i ) \exp\left\{ -\frac{1}{2 \sigma^2} \lVert \boldsymbol{x} - \boldsymbol x_i \rVert^2 \right\}.
\end{equation}
Here, $\mathcal{V}(\boldsymbol{x} \leftrightarrow \vect x_i )$ is a visibility indicator function between $\boldsymbol{x}$ and $\vect x_i$, \REPLACE{
defined as follows:
$$
\mathcal{V}(\boldsymbol{x} \leftrightarrow \boldsymbol{y} )\coloneqq\left\{
\begin{array}{ll}
0 & \textrm{ if }(1-t)\boldsymbol{x}+t\boldsymbol{y}\in\Domain\ \forall t\in[0,1]\\
1 & \textrm{ otherwise.}
\end{array}
\right.
$$}{
defined as $0$ if the line segment between $\vect{x}$ and $\vect x_i$ intersects the boundary and $1$ otherwise. %
}
This construction is illustrated visually in Figure~\ref{fig:rtfd}.

In practice, checking visibility distills down to a ray-mesh intersection query.
Efficient, robust algorithms support this query \cite[Chapter~6.8]{PBRT3e} on modern GPU hardware via the Optix library \cite{Parker10OptiX}. %
As a point of reference, a similar approach has been proposed to interpolate irradiance caches in real-time rendering \cite{Majercik2019Irradiance, halen2021global}.
To enable fast GPU radius queries within the hardware-accelerated ray tracing context, we truncate the kernel after a distance $r=3\sigma$ and employ a hash-grid data structure.
In Section~\ref{ss:optim}, we use a similar strategy to estimate smoothness energies while respecting the interior structure of the domain. %

\begin{remark}[Barycentric coordinates]
\citeauthor{Dodik:VBC:2023}'s work on barycentric coordinates [\citeyear{Dodik:VBC:2023}] provides a  point of contrast to our own.  The barycentric coordinates problem is closely linked to our Equation~\ref{eq:model}, with one additional ``Reproduction'' constraint.  This additional constraint, however, motivates their function parameterization, which is built by sampling simplices in the domain and combining their linear barycentric coordinate functions.  Thanks to the missing reproduction constraint, our parameterization is much simpler and does not scale cubically in the size of the domain.
\end{remark}

\subsection{Non-negativity and Partition of Unity}\label{ss:activation}

We can incorporate the non-negativity and partition of unity constraints into the final activation layer of $\Net_\theta$ using standard machine learning tools.
We can satisfy Equations~\ref{eq:nonneg}~and~\ref{eq:partition} by first applying an elementwise $\texttt{softplus}$ function \cite{dugas2000softplus} to the outputs of ${\Net_{\theta}}$---which maps them to the non-negative reals---and then normalizing such that they sum to one. Appendix~\ref{ap:funs} contains the full definitions of less common functions adopted in our work..%

\paragraph{Discussion} A more common choice in machine learning would be the exponential function instead of a \texttt{softplus}.
We found \texttt{softplus} to be preferable for several reasons.
A normalized exponential, also known as a soft maximum, amplifies the relative magnitude of the largest output.
However, we desire the opposite behavior as multiple control handles can exert influence over the same region.
More practically, an exponential activation has a larger span of possible gradient values, which yields instabilities during optimization.

\begin{figure}[ht!]
  \centering
  \includegraphics[width=\linewidth]{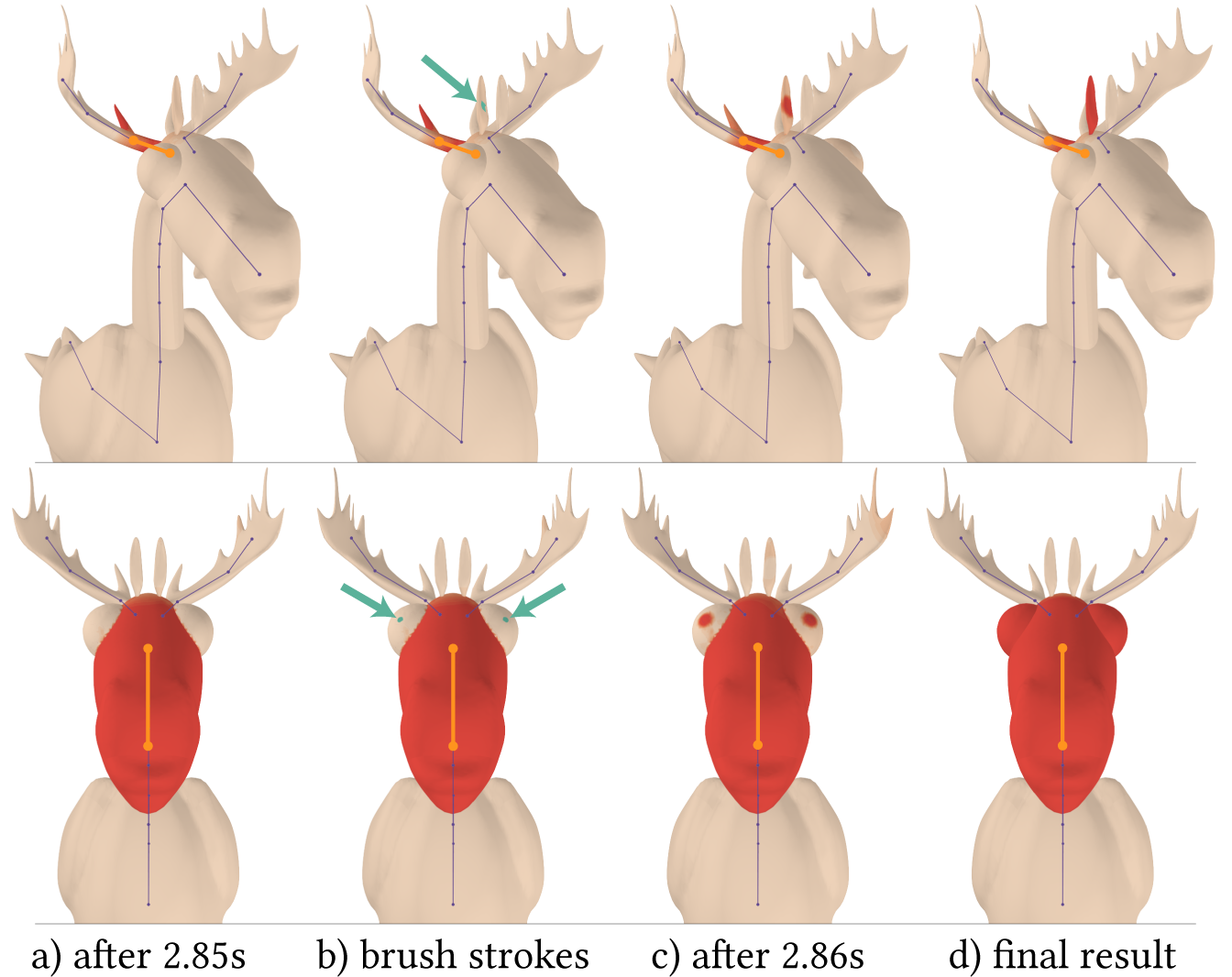}
  \caption{We combine the popular painting interface with our optimization method via Dirichlet boundary conditions. Our method quickly generates an initial solution (a). The ears and eyes of the \textsc{Goosemoose} mesh are disconnected as there is no path through the shape's interior that reaches them from any of the control handles. After inspection, the user can paint on skinning weights on a small subset of the boundary (b). The optimization then resumes, incorporating user-specified weights as Dirichlet boundary conditions (c, d).}
  \Description{a) mesh with an initial solution after $2.85$s of optimization, b) arrows pointing at brush stroke locations, c) weights after $2.85$s of optimization, d) final weights.}\label{fig:goosemoose}
\end{figure}

\subsection{Enforcing Boundary Conditions}\label{ss:bc}

It remains to enforce the Lagrange property \eqref{eq:lagrange}, itself a kind of Dirichlet boundary condition \eqref{eq:dirichlet}, and the Neumann boundary condition \eqref{eq:zeroneum}.
In physics-informed neural networks (PINNs), boundary conditions are commonly enforced \emph{weakly}, with a loss term \cite{lu2021hard,raissi2019physics}.
In this framework, it remains unclear how the interior and the boundary loss terms should be combined, leading to case-by-case parameter tuning and a brittle optimization procedure.
As a solution, recent work enforces such hard constraints by reparameterizing the function \cite{lu2021hard,Sukumar2022}.
We opt for this strategy due to its robustness. 
Our approach:
\begin{itemize}
    \item does not require a specific shape representation such as a signed distance field or restrict the topology or geometry of the boundary---we require only closest-point queries;
    \item only reparameterizes the function in an $\varepsilon$-neighborhood of the boundary, meaning that the majority of the computation remains unmodified; and
    \item only requires access to function values and not gradients, even when enforcing zero Neumann boundary conditions, reducing the necessary computation and avoiding the need for fourth-order derivatives.
\end{itemize}

\paragraph*{Lagrange property and Dirichlet boundary conditions}

The Lagrange property (Equation~\ref{eq:lagrange}) and Dirichlet boundary conditions (Equation~\ref{eq:dirichlet}) give hard constraints on the values of the skinning weights functions at certain locations.  The Lagrange property is automatically enforced at the bones, while the Dirichlet conditions are optionally painted on.  We describe how we implement the Lagrange property below; Dirichlet conditions are implemented in a nearly identical fashion.

We begin by restating Equation~\ref{eq:lagrange} in terms of a control handle indicator function, ${\vect{e} : \Domain \rightarrow \{0, 1\}^K}$. We define $\vect{e}$ such that, for all $\DomainSample$ that lay on handle $\Handle_i$, the $i$\textsuperscript{th} element of $\vect{e}(\DomainSample)$ equals $1$ and all other elements equal $0$:
\begin{equation}
    e_i(\DomainSample) \coloneq %
    \begin{cases}
        1 & \text{if $\DomainSample \in \Handle_i$},\\
        0 & \text{otherwise}.
    \end{cases}
\end{equation}
It follows that the Lagrange condition can be satisfied via
\begin{equation}~\label{eq:sharpbc}
  \baryv(\DomainSample) = \vect{e}(\DomainSample) +
  \left({1 - \sum_{i=1}^{K} e_i \left(\DomainSample\right)} \right)\,\Net_\theta(\DomainSample)
  .
\end{equation}
In general, this transformation makes $\Net_\theta$ discontinuous at the handles. Therefore, to enforce this boundary condition, we construct a \emph{mollified} version of $\vect e$---denoted as $\widetilde{\vect{e}}$---that equals $\vect e$ on the control handles and quickly falls off to zero as we move away.
As such, we are essentially interpolating between $\Net_\theta(\DomainSample)$ and $\vect{e}\left(\DomainSample\right)$ as $\DomainSample$ approaches a control handle inside cylindrical $\varepsilon$-neighborhoods of bones and spherical $\varepsilon$-neighborhoods of point handles.
\begin{equation}~\label{eq:smoothbc}
    \baryv(\DomainSample) 
\coloneqq
    \widetilde{\vect{e}}\left(\DomainSample\right) + %
    \left({1 - \sum_{i=1}^{K} \widetilde{e}_i\left(\DomainSample\right)} \right)\,\Net_\theta(\DomainSample).%
\end{equation}
We smooth out \eqref{eq:sharpbc} by replacing the binary $0$ or $1$ decision inside of $e_i$ with a so-called smooth \emph{bump function}, $w \colon [0, \varepsilon] \to [0, 1]$, ensuring $w(0) = 1$, $w(\varepsilon) = 0$, and $w'(0) = 0$.
The exact construction of $\widetilde{e}_i$ and the choice of $w$ are described in Appendix~\ref{ap:funs}.

\begin{figure}[ht!]
  \centering
  \includegraphics[width=\linewidth]{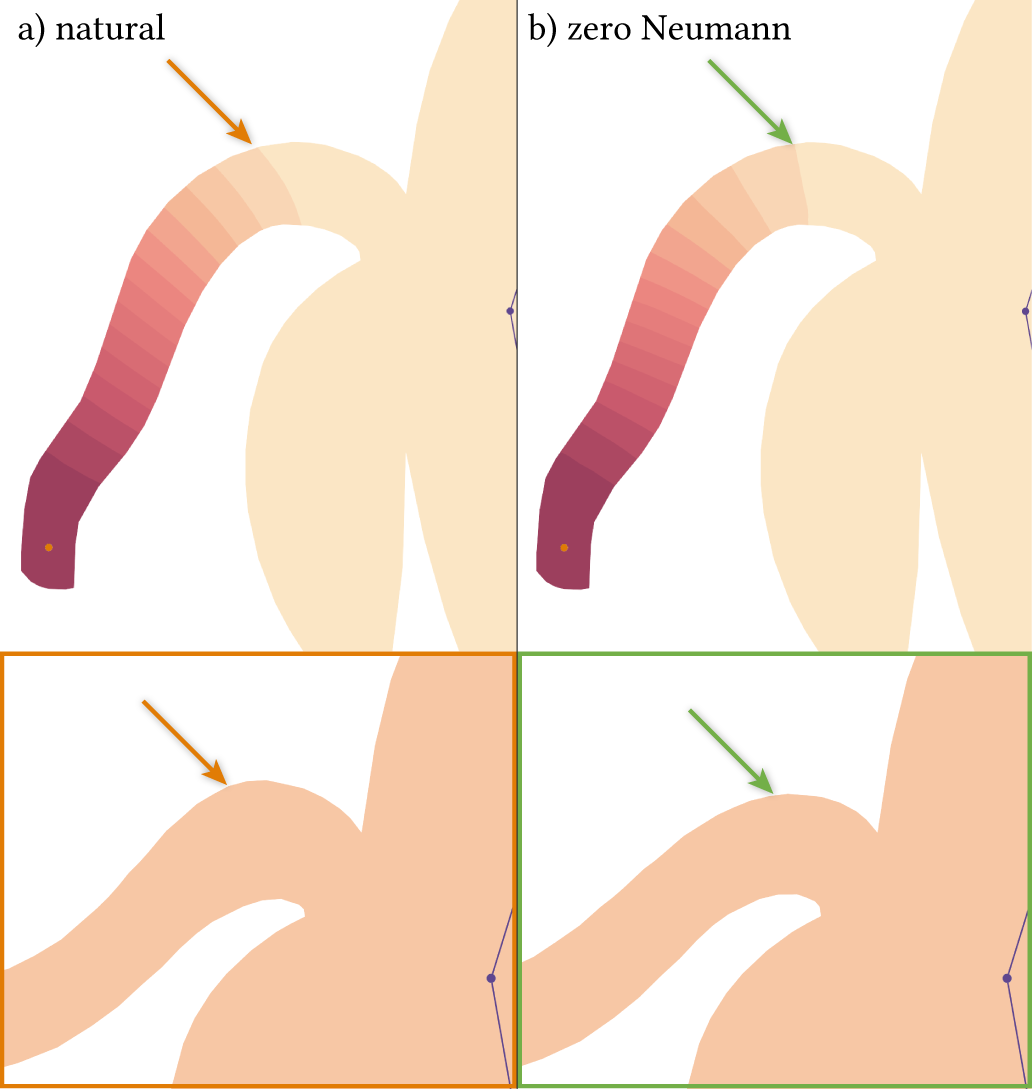}
  \caption{The user can optionally enable zero Neumann boundary conditions, making the level sets of the weights orthogonal to the boundary. In this example, zero Neumann weights (top) make it so that the control handle symmetrically deforms its region of influence (bottom).}
  \Description{2D mesh with weights with and without zero Neumann conditions.}\label{fig:bananeuman}
\end{figure}

\paragraph*{Zero Neumann boundary conditions.}\label{sec:neumann} 
To match the implementation of~\citet{BBW:2011}, we optionally modify $\baryv$ near the boundary to enforce the zero Neumann boundary condition (Equation~\ref{eq:zeroneum}). Its effects can be seen in Figure~\ref{fig:bananeuman}.

To evaluate $\baryv$ with zero Neumann conditions at $\pos \in \Domain$, we first find the closest boundary point $\bdr \in \Boundary$.
We will refer to the distance between $\pos$ and $\bdr$ as $t$ and to the surface normal at $\bdr$ as $\normal_{\bdr}$.
Our goal is to edit $\baryv$ to enforce that, as ${t \rightarrow 0}$, we have $(\baryv(\pos)-\baryv(\bdr))/t\to\vect0$.

\begin{figure}[h!]
  \centering
  \includegraphics[width=\linewidth]{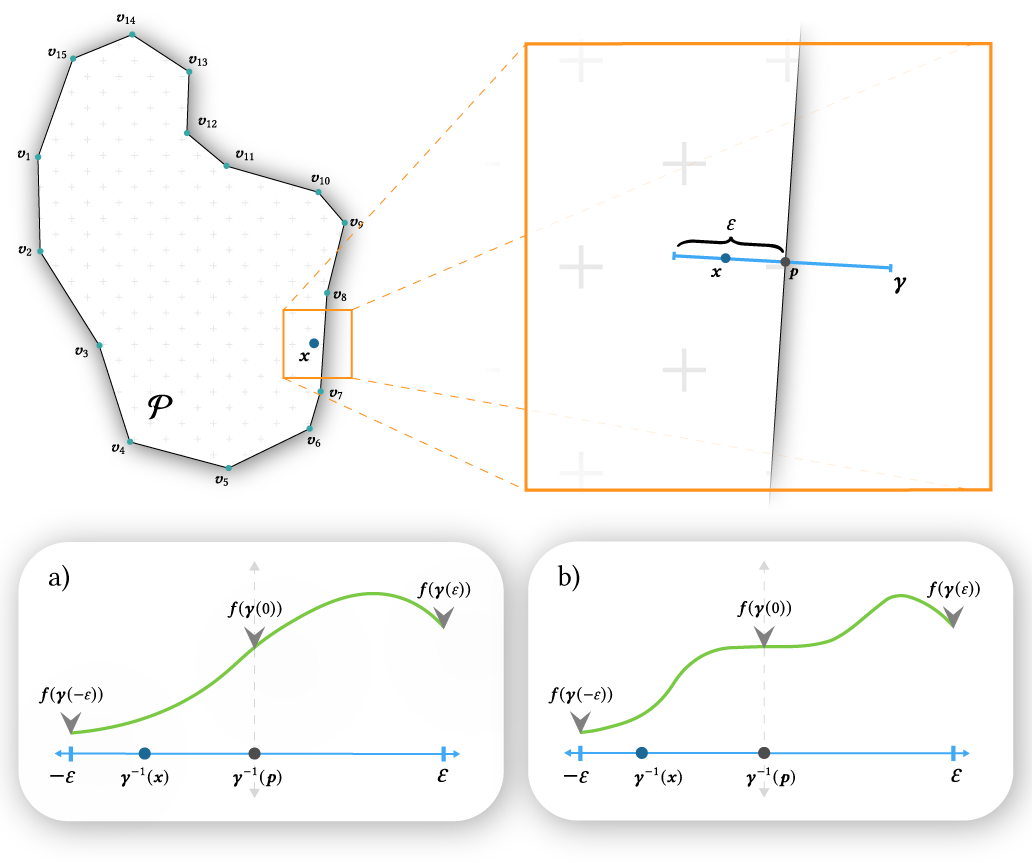}
  \caption{An illustration of our zero Neumann reparameterization within an $\varepsilon$-neighborhood of the boundary. }
  \Description{An illustration of our zero Neumann reparameterization within an epsilon-neighborhood of the boundary.}\label{fig:neumanbc}
\end{figure}

As there are many ways to design this procedure, %
we opt to preserve as much of the original function as possible, i.e.,  $\baryv$ should remain unmodified  for all $\pos$ outside an $\varepsilon$ neighborhood of $\Boundary$. 
We begin by defining a parametric line, ${\fline: [-\varepsilon, \varepsilon] \NarrowArrow \Domain}$, that connects $\pos$ and $\bdr$, satisfying 
${\fline(0) = \bdr}$, ${\fline(t) = \pos}$, and $\fline' \equiv \normal_{\bdr}$.
This allows us to rewrite the boundary condition as
\begin{equation}
    \frac\Diff{\Diff t}
    \baryv(\fline(t))
    \Big\vert_{t = 0} = 
     \normal_{\bdr}\cdot \nabla \baryv(\bdr)%
    = 0.
\end{equation}
Denoting the unmodified function as $\hat\baryv$, we can write %
\begin{equation}
  \baryv(\pos) =
  \hat\baryv(\pos) + w\left(t%
  \right) \left({\hat\baryv(\bdr) - \hat\baryv(\pos)}\right),
\end{equation}
where we use $w \colon [0, \varepsilon] \NarrowArrow [0, 1]$ \REPLACE{is an interpolant function} with $w(0) = 1$, $w(\varepsilon) = 0$, $w'(0) = 0$, as defined previously and in Appendix~\ref{ap:funs}.
This formulation is illustrated in Fig.~\ref{fig:neumanbc}.
By construction, $\baryv$ satisfies the following property:
\begin{proposition}
$\baryv$
satisfies zero Neumann conditions for any smooth $\hat\baryv$ on the interiors of the boundary facets of $\Domain$.
\end{proposition}

\begin{proof}
We need to check for $\normal_{\bdr}\cdot\nabla\baryv(\bdr)=0.$ %
By the chain rule,
\begin{align*}
\normal_{\bdr}\cdot\nabla\baryv(\bdr)
&= \normal_{\bdr} \!\cdot\! (\nabla \hat\baryv(\bdr) 
\!+\!  w'\left( 0 \right) \underbrace{\left({\hat\baryv(\bdr) \!-\! \hat\baryv(\bdr)}\right)}_{=\vect0} \!-\! \underbrace{w \left( 0 \right) }_{=1} {\nabla \hat\baryv(\bdr)} \Big)  \\
&= \normal_{\bdr} \cdot \Big(\nabla {\hat\baryv(\bdr)} - \nabla \hat\baryv(\bdr) \Big) = 0.
\end{align*}
In the first equality, note that the term next to $w(0)$ contains only a single derivative of $\hat\baryv$; the other term vanishes since we are projecting onto the flat boundary facets of $\Domain$.
\end{proof}

\noindent The proposition above has to be stated carefully because polygonal choices of $\Domain$ necessarily have sharp corners, where Neumann conditions are ill defined.  Empirically, we find that behavior of our model near these corners is still reasonable and converges to the proper global boundary conditions as the boundary is refined.

\section{Optimization}\label{ss:optim}

With our representation of skinning weight functions in place, we proceed to describe our method for computing and optimizing the biharmonic energy. Our approach uses a randomized estimator compatible with stochastic gradient-based optimization techniques.

\subsection{Estimation and Optimization of Biharmonic Energy}\label{ss:est}

In each iteration of the optimization, we uniformly randomly sample $M$ samples $\{\vect y_j\}_{j=1}^M$ in the interior of the domain $\Domain$, used as quadrature points to compute our approximation of the biharmonic energy objective.
If a mesh of the interior is available, we use it for random sampling of the interior; this mesh is used only for sampling and does \emph{not} have to satisfy the quality conditions that are typically necessary for FEM.
If a mesh is unavailable, we randomly sample the ambient space and use the generalized winding number \cite{Jacobson2013Winding} to reject samples below a threshold, as described below in  Section~\ref{ss:gwn}.
It is important to sample points in the interior of the domain and not just on the boundary surface, since the integral in (\ref{eq:model}) is formulated on the \emph{volume}.

Given our set of samples, we use a Monte Carlo estimate of the bilaplacian energy in Equation~\ref{eq:bbw}:
\begin{equation}\label{eq:energysum}
     \sum_{i=1}^{K} \int_\Domain \vert \Laplacian \bary_i(\vect{y}) \vert^2 \Diff V(\vect{y}) %
     \approx \frac{V(\Domain) }{M} \sum_{i=1}^{K} \sum_{j=1}^M \vert \hat{\Laplacian} \bary_i(\vect y_j) \vert^2,
\end{equation}
where $\hat \Laplacian$ is a visibility-aware finite-difference Laplacian estimator defined  in Section~\ref{ss:fd}.  

We optimize this objective using the Adam optimizer \cite{kingma2014adam}.  A new set of samples $\mathbf y_j$ is drawn in each iteration of Adam.  Note \citet{wang2021fast} also find Adam to be an effective optimizer for skinning weights problems. %
Please see Appendix~\ref{s:impl} for more implementation details.

\subsection{Generalized-Winding Numbers Bilaplacian Energy}\label{ss:gwn}

We are able to apply our method to non-watertight geometries through an extension that relies on robust methods popular in other parts of geometry processing.

In particular, we rely on generalized winding numbers \cite{Jacobson2013Winding, barill2018fast} to perform robust inside-outside segmentation.
For a watertight shape, the generalized winding number is a function $\gwn: \R^d \rightarrow \R$, which equals $1$ for all points inside the shape and $-1$ for those outside of the shape.
For an open shape, the generalized winding number smoothly varies from $1$ to $-1$ as the percentage of the shape's faces surrounding a point decreases.

We define two thresholds, $\gwn_{l}$ and $\gwn_{h}$.
Anything below ${\gwn_{l}}$ is considered completely outside the shape, and anything above ${\gwn_{h}}$ is completely inside. Our implementation uses ${\gwn_{l} = 0.1}$ and ${\gwn_{h} = 0.25}$.

We use $\gwn_{l}$ to reject points both when sampling the point cloud $\{\vect{x}_i\}$, as well as the optimization samples $\{\vect{y}_i\}$.
For all remaining points, %
we clip their generalized winding numbers above $\gwn_{h}$ compute a weighted biharmonic energy:
\begin{equation}\label{eq:energysum}
     \frac{V(\Domain) }{M} \sum_{i=1}^{K} \sum_{j=1}^M \frac{\gwn(\vect y_j) - \gwn_{l}}{\gwn_{h} - \gwn_{l}} \vert \hat{\Laplacian} \bary_i(\vect y_j) \vert^2.
\end{equation}

\begin{figure}[h]
  \centering
  \includegraphics[width=\linewidth]{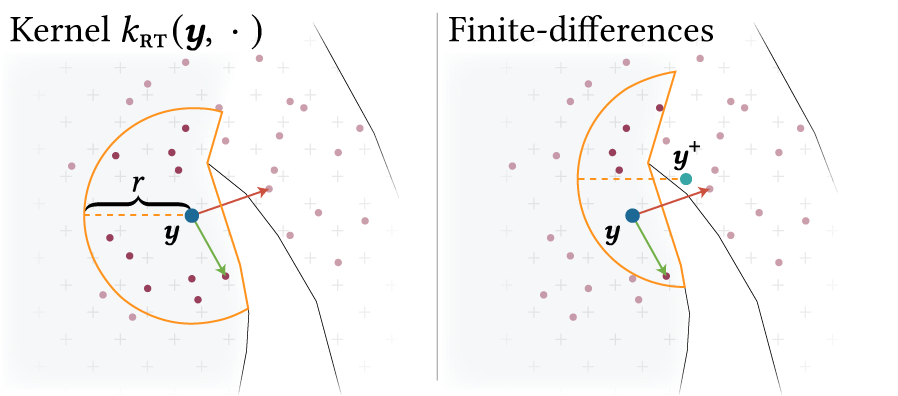}
  \caption{The figure on the left visualizes our ray-tracing kernel $k_\textsc{rt}$. For an evaluation point $\vect{y}$, we look up all points within a radius $r$, and include into our estimate points which are visible from $\vect{y}$. Here we visualize a visible point with a green arrow pointing to it, and an occluded point with a red arrow. The figure on the right illustrates our visibility-aware finite-difference scheme. Our construction ensures that the kernel estimate at $\vect{y}^{+}$ can only include points which are in a star-shaped neighborhood of $\vect{y}$.}
  \Description{Illustration of the ray tracing kernel and finite differences in star shaped neighborhoods.}\label{fig:rtfd}
\end{figure}

\subsection{Finite Differences}\label{ss:fd}

Previous work highlights problems in relying on automatic differentiation to compute higher-order derivatives of function representations like ours \cite{li2023neuralangelo, chetan2023accurate}; in particular, differential quantities obtained through automatic differentiation tend to be noisy, and using them for optimization results in visible artifacts \cite{chetan2023accurate}.
On the other hand, a na\"ive finite-difference estimator can yield bleeding artifacts similar to those in Figure~\ref{fig:rt}.  Here, we suggest a stochastic estimator of the Laplacian that reduces computation time and avoids bleeding.

A standard finite-difference estimator of the Laplacian, $\Laplacian_{h}$ can be defined as
\begin{equation}\label{eq:standardlaplacian}
    \Laplacian_{h} f(\mathbf y) = \frac{-2 d f(\mathbf y) + \sum_{i=1}^{d} (f(\mathbf y + h \vect{e}_i) + f(\mathbf y - h \vect{e}_i))}{h^2},
\end{equation}
where $h$ is the finite-difference step size, and $\vect{e}_i$ represents the $i$\textsuperscript{th} basis vector in $\R^d$.
This well-known finite-difference Laplacian requires a linear number $2d + 1$ of computations in the dimensionality $d$ of the ambient space; this means a single term on the right-hand side of \eqref{eq:energysum} would require $7$ evaluations of $\bary_i$ in $3$D ($d=3$) and $5$ evaluations in $2$D ($d=2$).

Instead, since our evaluation of the objective function is stochastic anyway, we use a modified stochastic Laplacian estimator:
\begin{equation}\label{eq:randomlaplacian}
    \hat \Laplacian_{h} f(\mathbf y) = \frac{-2 f(\mathbf y) + f(\mathbf y + h \mathbf{v}) + f(\mathbf y - h \mathbf{v})}{h^2},
\end{equation}
where $\mathbf{v}$ is uniformly drawn from the unit sphere $S^{d-1}$; we draw a different $\mathbf{v}$ for each term in \eqref{eq:energysum}. The expectation of this expression over $\mathbf{v}\in S^{d-1}$ converges to $\Delta f$ as $h\to0$. %
Our expression \eqref{eq:randomlaplacian} simultaneously reduces the number of function evaluations to a constant independent of dimension, and it removes the axis alignment bias of the standard estimator \eqref{eq:standardlaplacian}.

\begin{remark}[bias]
There are two sources of minor bias in our estimate \eqref{eq:energysum} relative to the true biharmonic energy on the left-hand side.  First, we use a positive step size $h>0$ rather than limiting $h\to0$.  Second, by Jensen's inequality, squaring \eqref{eq:randomlaplacian} leads to a slight overestimate of the true objective on average; this bias is also present in the Dirichlet energy of \citet{Dodik:VBC:2023}.  We do not find either source of bias to be significant and leave derivation of practical fully-debiased estimates of the biharmonic energy to future work.
\end{remark}

We now address bleeding in the evaluation of the smoothness objective.  Take $\vect{y}^\pm\coloneqq \vect{y}\pm h\mathbf v$ to be a displaced sample used in evaluation of \eqref{eq:randomlaplacian}.  When evaluating $\Net_{\theta}(\vect{y}^\pm)$, we simply zero out the kernel $k$ for those points $\vect{x}_i$ that are not visible from $\vect{y}$.
In other words, we ensure that all $\vect{x}_i$ used for the evaluation of the Laplacian at $\vect y$ are within a star-shaped neighborhood of $\vect y$ (see Fig.~\ref{fig:rtfd}).

\subsection{Multiscale Optimization}\label{ss:init}

Using a gradient descent method for optimizing smoothness energies incurs a trade-off. 
If we use a finite-difference step that is too small, we will have to wait longer for the boundary conditions to propagate inward.
Since our formulation is nonconvex, this issue also can lead to local minima.

In mesh-based pipelines, these difficulties are resolved via preconditioning, either by using Sobolev gradients or through higher-order methods \cite{wang2021fast}.
In our scenario, however, it is not clear how to precondition the optimization of the biharmonic energy over a smooth geometric field that is nonconvex in its parameters.
Therefore, we introduce a simple yet effective alternative to preconditioning that results in high-resolution solutions while reducing the number of iterations for convergence.

Inspired by multigrid methods and prior work on optimizing differential energies in neural fields \cite{li2023neuralangelo}, we found that a multiscale approach offers an effective  solution.
Specifically, we initiate our method by randomly sampling $\mathcal{X}$.
We let the method optimize for a number of iterations before doubling the number of samples in $\mathcal{X}$.
To initialize the new point cloud, we use $\Net_\theta$ to interpolate existing values $\vect{f}_i$ onto the new points.
Additionally, we also interpolate Adam's gradient moment estimates~\cite{kingma2014adam} in a similar fashion.
As the radius of our kernel $k$ and the finite-differences step size are related to the density of the point cloud (see Figure~\ref{fig:ablation}), they shrink as we up-sample the point-cloud.

The upsampling generally increases the variance in the evaluation of the biharmonic energy as the interpolation kernel and finite-difference estimator progressively do less and less smoothing (see Figures~\ref{fig:multiscale}~and~\ref{fig:msloss}).

To stabilize the optimization, we progressively decrease the learning rate every time we upsample the underlying representation (see Appendix~\ref{s:impl} for details).
We demonstrate the practical effect of this strategy qualitatively in Figure~\ref{fig:multiscale} and quantitively in Figure~\ref{fig:msloss}.

\begin{figure}[h]
  \centering
  \includegraphics[width=\linewidth]{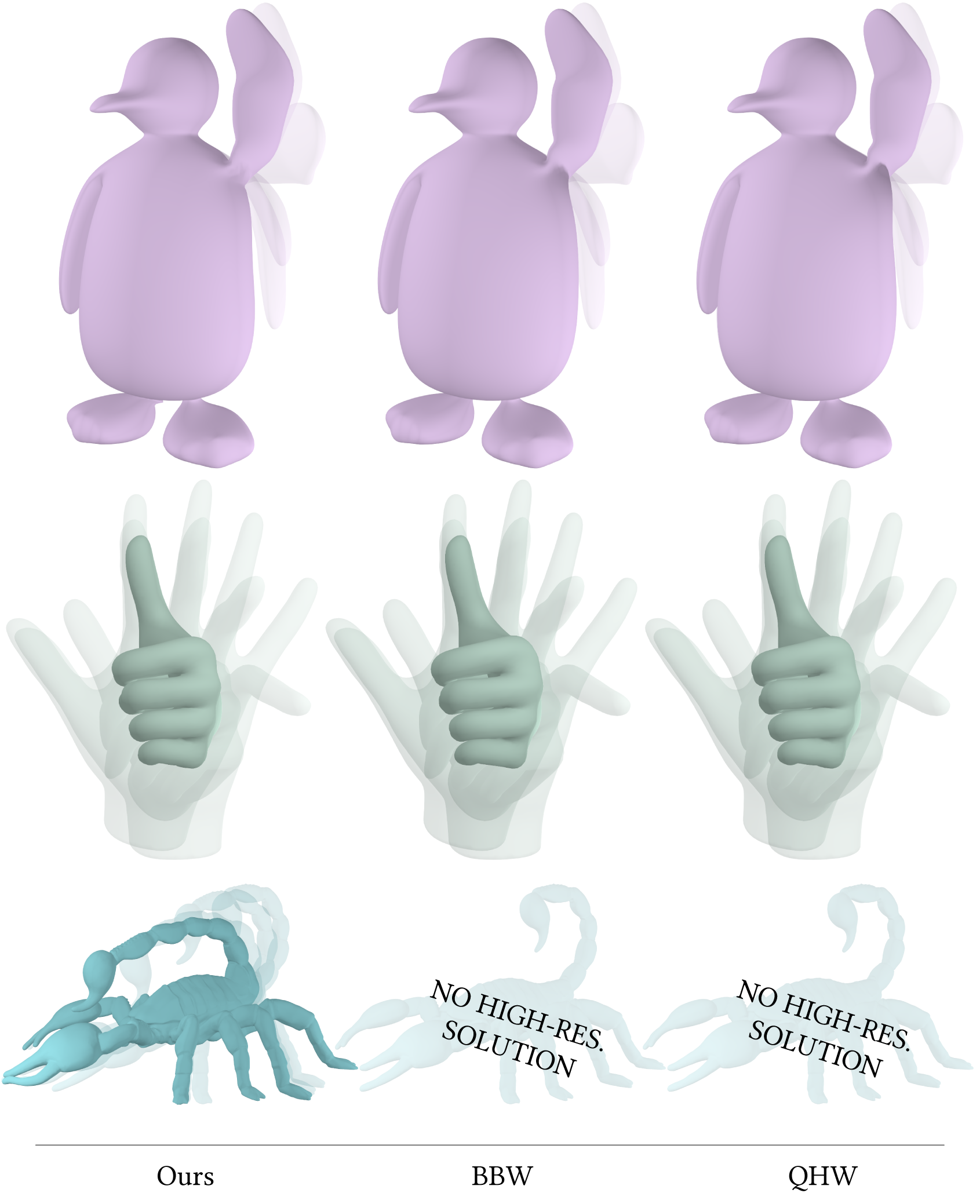}
  \caption{We compare the deformations produced by our skinning weights with previous volumetric skinning weights methods: BBW \cite{Jacobson2013Winding} and QHW \cite{wang2021fast}. As expected, all methods produce similar results on simple geometries such as the \Penguin{} and \Hand{} meshes. However, robust tetrahedralization software is not able to produce a high-resolution solution for the more complicated \Scorpion{} mesh.}
  \Description{Animation comparisons on three different meshes.}\label{fig:companim}
\end{figure}

\begin{figure*}[ht!]
  \centering
  \includegraphics[width=\linewidth]{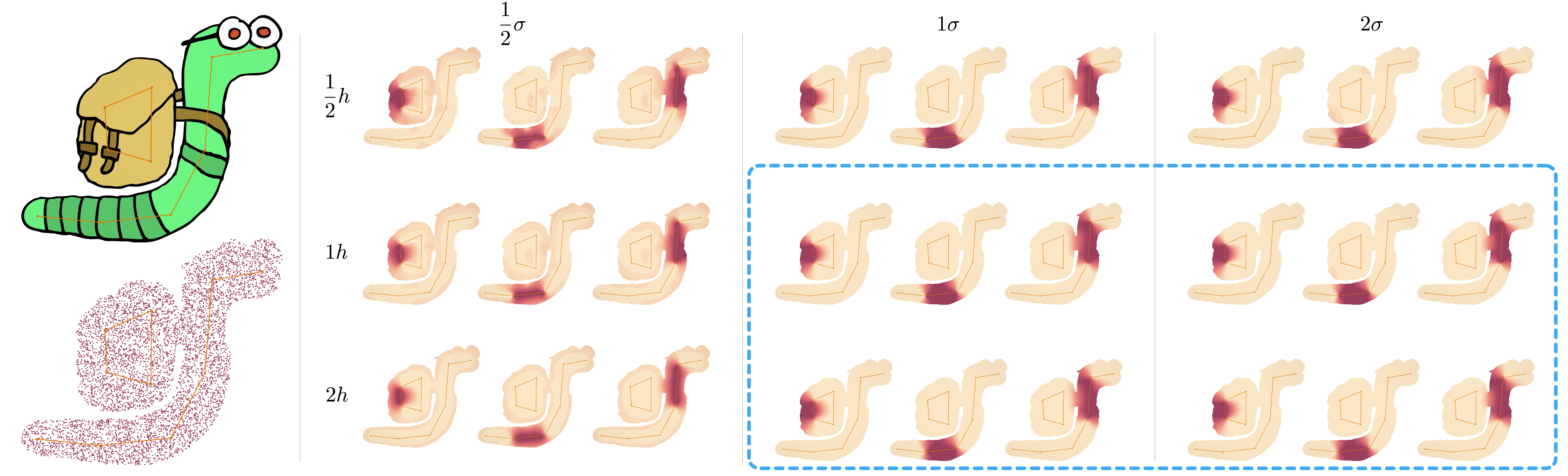}
  \caption{Results of our ablation study. On the top, we show the effect of varying the kernel parameter $\sigma$, as well as the finite-difference step size, $h$. Making $\sigma$ or $h$ too small compared to the median spacing of points makes optimizing differential quantities difficult and leads to wrong results. Doubling the size of $\sigma$ or $h$ produces a similar result and would be viable choices of parameters, but it does unnecessarily increase the kernel look-up time.}
  \Description{Ablation study on a 2D mesh comparing the effect of kernel size and finite-difference step size.}\label{fig:ablation}
\end{figure*}

\section{Results}\label{s:results}

We ran all of our experiments on a machine with an Intel i9-13900 CPU, $32$ GB of memory, and an Nvidia GeForce RTX 4090.
See Appendix~\ref{s:impl} for implementation details, including used libraries and hyper-parameter choices.
The comparison results for bounded biharmonic weights (BBW)~\cite{BBW:2011} were generated in Matlab using the \texttt{gptoolbox} library~\cite{gptoolbox}, and for quasi-harmonic weights (QHW) \cite{wang2021fast} using the authors' implementation.
All of our deformations were produced using dual quaternion skinning \cite{Kavan08DQS}.
We will release the code upon acceptance.

To the best of our knowledge, there are currently no large-scale datasets or benchmarks for robust skinning weight computation.
To evaluate our algorithm, we curated a sample of $3$D meshes and manually created the skeletons.
Most of the meshes are based on diverse and realistic scenarios for animation; $5$ are clean and well-behaved, $3$ are visually clean yet happen to break previous methods (\textsc{Scorpion}, \textsc{Gear}, \textsc{Cow}), and $4$ consist of open and self-intersecting meshes with poor quality elements that a user might nonetheless plausibly wish to animate (\Beaver{}, \Piggybank{}, \Bunny{}, \textsc{Goosemoose}).
Lastly, \textsc{Scorpion Rand.}, \textsc{Gear}, and the meshes in Figure~\ref{fig:open} represent extreme stress-test scenarios that are likewise handled well by our algorithm.

We first validate the properties of our method and ablate parts of our algorithm and our parameter choices in Section~\ref{ss:valid}.
To provide evidence for robustness, Section~\ref{ss:comp} offers a qualitative and quantitative comparison against previous work that solves for bounded biharmonic weights.
We then examine the influence of different boundary conditions in Section~\ref{ss:bcres}.
Finally, we discuss limitations and future work in Section~\ref{ss:limit}.

\subsection{Validation and Ablation}\label{ss:valid}

We first validate that our weights look correct and ablate the algorithm's hyperparameters.
We attempt to tetrahedralize all $3$D meshes before comparing them with BBW \cite{BBW:2011} and QHW \cite{wang2021fast} in Table~\ref{tbl:qant} and Figures~\ref{fig:timings},~\ref{fig:companim},~and~\ref{fig:comparison}.

Figure~\ref{fig:comparison} compares our skinning weights with weights computed using bounded biharmonic weights (BBW)~\cite{BBW:2011}, as well as those computed using quasi-harmonic weights (QHW)~\cite{wang2021fast}.
As expected, all methods produce qualitatively similar weights functions, but the setting is quite different: ours can work with a boundary mesh/polygon, while BBW/QHW require filling the domain with elements by tet-/tri-meshing.
Figure~\ref{fig:companim} compares the animations produced by our method to those produced by BBW and QHW.
Again, the results are nearly indistinguishable on simple meshes.
The major difference is that our method does not require tetrahedralization.%

\begin{figure}[hb!]
  \centering
  \includegraphics[width=\linewidth]{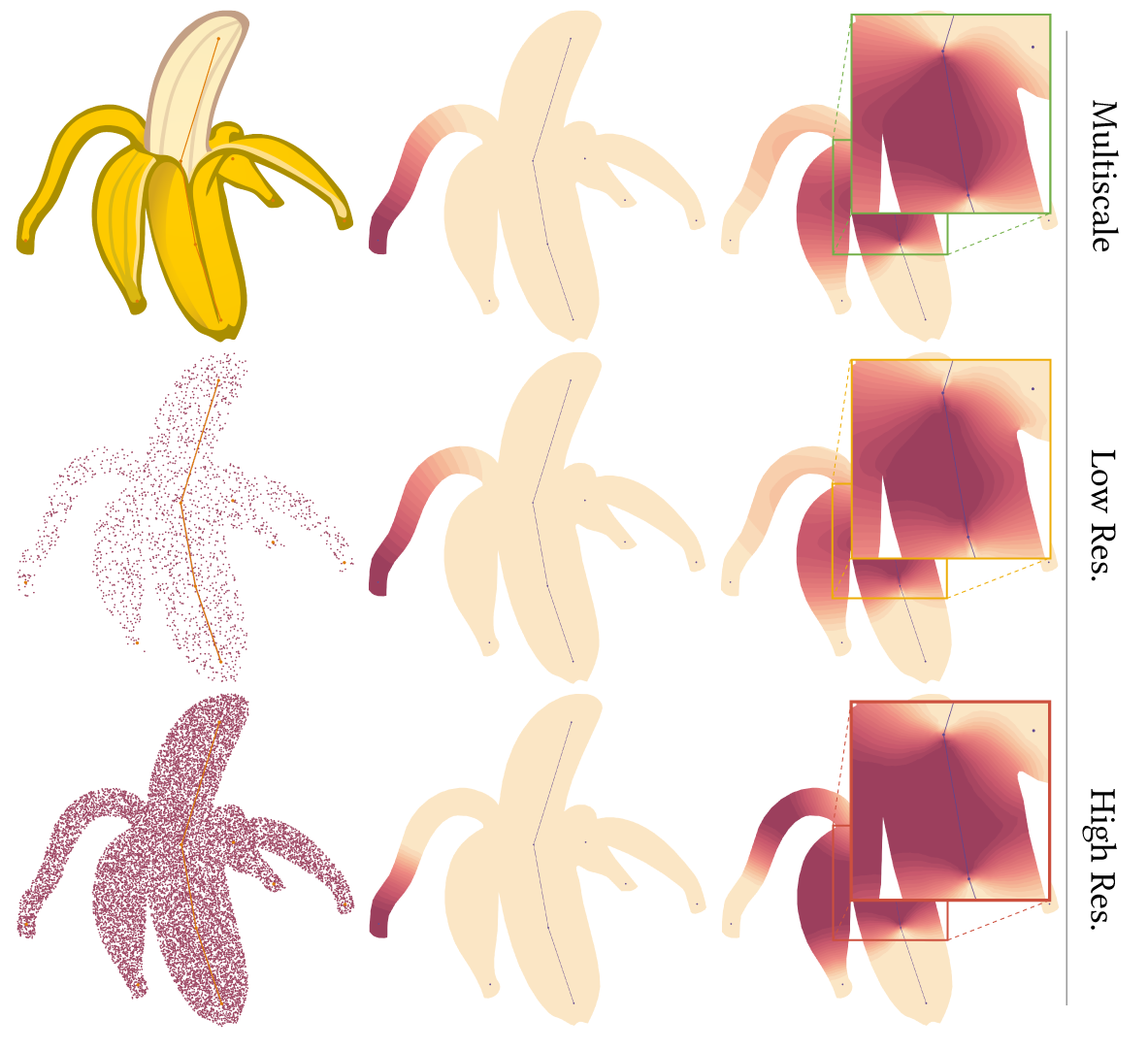}
  \caption{Final results of our multiscale optimization strategy (top row) on \Banana{} (top left). When optimizing with a coarse point cloud of $2^{11}$ samples, the optimization converges quickly, but the function is overly smooth near the control handles (middle row). Na\"ively increasing the resolution to $2^{14}$ points results in slow convergence and possible local minima (bottom row).}
  \Description{Qualitative comparison of the our multiscale strategy with a fully low-res or a fully high-res optimization strategy on a 2D mesh.}\label{fig:multiscale}
\end{figure}

\paragraph*{Ablation} Our method includes some parameters, the most important of which are the number of points in $\mathcal{X}$, the size of $\sigma$, and the size of the finite-difference step $h$.
To determine suitable values, we ran a series of ablation studies.

Figure~\ref{fig:ablation} demonstrates the effects of varying $\sigma$ and $h$.
There is an inherent tension regarding the size of $\sigma$.
On one hand, we want $\sigma$ to be as small as possible to accelerate radius queries.
On the other, too small of a $\sigma$ results in what is effectively a $1$-nearest-neighbor estimator, which makes estimating differential quantities difficult.
As a compromise, we found the following to be a useful heuristic: We set $\sigma$ to the median distance to the $9$\textsuperscript{th} nearest neighbor.
As shown in our ablation, this choice of $\sigma$ makes it large enough to avoid any issues associated with a degenerate kernel estimator; making it significantly larger produces correct, if slightly blurry, results but offers no benefit while increasing the runtime.

\begin{figure}[hb!]
  \centering
  \includegraphics[width=\linewidth]{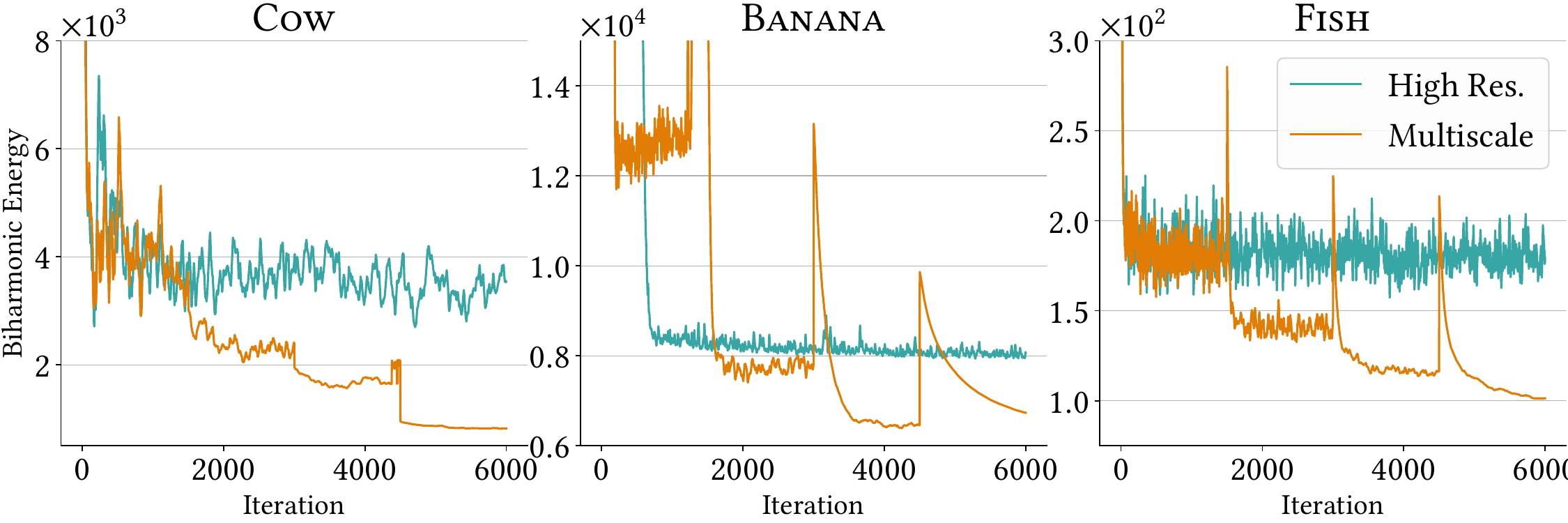}
  \caption{Our multiscale optimization results in lower biharmonic energy compared to a fixed high-resolution sampling of $\mathcal{X}$.}
  \Description{}\label{fig:msloss}
\end{figure}

\paragraph*{Multiscale optimization}
The multiscale optimization strategy is an important component of our method; we demonstrate its benefits by comparing it to fixed-size $\mathcal{X}$.

We qualitatively compare different fixed sizes of of $\mathcal{X}$ to our strategy in Figure~\ref{fig:multiscale}.
Too small of a point cloud results in an overly blurry representation near the handles, whereas too large of a point cloud results in a higher loss, slower convergence, and possible local minima.
In comparison, our multiscale optimization combines the benefits of both.

\begin{table*}[ht!]
\begin{NiceTabularX}{\textwidth}{Xcrrrrrrrr}
\CodeBefore
    \rowcolors{2}{gray!10}{}[restart,respect-blocks]
\Body
\toprule
&\thead[c]{Meshing \\ type} &\thead[c]{Meshing (s)} &\thead[c]{Interior\\ points} &\thead[c]{BBW \\ Optim. (s)} &\thead[c]{QHW \\ Optim. (s)} &\thead[c]{BBW \\ Total (s)} &\thead[c]{QHW \\ Total (s)} &\thead[c]{\textbf{Ours} \\ \textbf{Natural BC (s)}} &\thead[c]{\textbf{Ours} \\ \textbf{Neumann BC (s)}} \\\midrule
\textsc{Hand} &tetgen &7.64 &67341 &250.07 &29.29 &257.7 &\textbf{36.92} &59.79 &82.37 \\
\textsc{Penguin} &tetgen &8.05 &67440 &212.57 &25.19 &220.62 &\textbf{33.24} &58.67 &85.54 \\
\textsc{Fish} &tetgen &2.26 &\cellcolor[HTML]{ffe599}18875 &71.17 &7.86 &73.43 &\textbf{10.13} &56.86 &80.68 \\
\textsc{Mushroom} &tetgen &7.8 &65738 &270.5 &28.1 &278.29 &\textbf{35.9} &59.95 &82.45 \\
\textsc{Shibainu} &tetgen &7.79 &65596 &391.66 &44.22 &399.46 &\textbf{52.01} &76 &102.45 \\
\midrule
\multirow{2}{*}{\Block[l]{2-1}{\textsc{Scorpion}}} &default &26.19 &\cellcolor[HTML]{ffe599}54963 &535.47 &31.03 &561.66 &57.22 &\multirow{2}{*}{\textbf{46.31}} &\multirow{2}{*}{81.91} \\
&accurate &\cellcolor[HTML]{ea9999}Failed &\cellcolor[HTML]{ea9999}Failed &\cellcolor[HTML]{ea9999}Failed &\cellcolor[HTML]{ea9999}Failed &\cellcolor[HTML]{ea9999}Failed &\cellcolor[HTML]{ea9999}Failed & & \\
\multirow{2}{*}{\Block[l]{2-1}{\textsc{Scorpion (Rand.)}}} &default &\cellcolor[HTML]{ea9999}Failed &\cellcolor[HTML]{ea9999}Failed &\cellcolor[HTML]{ea9999}Failed &\cellcolor[HTML]{ea9999}Failed &\cellcolor[HTML]{ea9999}Failed &\cellcolor[HTML]{ea9999}Failed &\multirow{2}{*}{\textbf{48.23}} &\multirow{2}{*}{82.27} \\
&accurate &\cellcolor[HTML]{ea9999}Failed &\cellcolor[HTML]{ea9999}Failed &\cellcolor[HTML]{ea9999}Failed &\cellcolor[HTML]{ea9999}Failed &\cellcolor[HTML]{ea9999}Failed &\cellcolor[HTML]{ea9999}Failed & & \\
\multirow{2}{*}{\Block[l]{2-1}{\textsc{Gear}}} &default &27.53 &\cellcolor[HTML]{ffe599}43128 &117.25 &15.57 &144.78 &43.11 &\multirow{2}{*}{\textbf{30.48}} &\multirow{2}{*}{50.65} \\
&accurate &5974.59 &63072 &440.33 &63.52 &6414.92 &6038.1 & & \\
\multirow{2}{*}{\Block[l]{2-1}{\textsc{Beaver}}} &default &\cellcolor[HTML]{ea9999}Failed &\cellcolor[HTML]{ea9999}Failed &\cellcolor[HTML]{ea9999}Failed &\cellcolor[HTML]{ea9999}Failed &\cellcolor[HTML]{ea9999}Failed &\cellcolor[HTML]{ea9999}Failed &\multirow{2}{*}{\textbf{38.97}} &\multirow{2}{*}{66.93} \\
&accurate &\cellcolor[HTML]{ea9999}Failed &\cellcolor[HTML]{ea9999}Failed &\cellcolor[HTML]{ea9999}Failed &\cellcolor[HTML]{ea9999}Failed &\cellcolor[HTML]{ea9999}Failed &\cellcolor[HTML]{ea9999}Failed & & \\
\multirow{2}{*}{\Block[l]{2-1}{\textsc{Piggybank}}} &default &54.12 &65981 &132.47 &22.34 &186.59 &76.46 &\multirow{2}{*}{\textbf{45.99}} &\multirow{2}{*}{69.91} \\
&accurate &269.73 &66502 &221.4 &40.61 &491.13 &310.34 & & \\
\multirow{2}{*}{\Block[l]{2-1}{\textsc{Bunny}}} &default &361.74 &64382 &737.58 &73.28 &1099.32 &435.01 &\multirow{2}{*}{\textbf{59.58}} &\multirow{2}{*}{91.16} \\
&accurate &\cellcolor[HTML]{ea9999}Failed &\cellcolor[HTML]{ea9999}Failed &\cellcolor[HTML]{ea9999}Failed &\cellcolor[HTML]{ea9999}Failed &\cellcolor[HTML]{ea9999}Failed &\cellcolor[HTML]{ea9999}Failed & & \\
\multirow{2}{*}{\Block[l]{2-1}{\textsc{Cow}}} &default &50.64 &61500 &364.03 &39.03 &414.67 &89.67 &\multirow{2}{*}{\textbf{40.00}} &\multirow{2}{*}{84.54} \\
&accurate &1593.12 &63894 &2260.49 &\cellcolor[HTML]{ea9999}Failed &3853.6 &1592.12 & & \\
\multirow{2}{*}{\Block[l]{2-1}{\textsc{Goosemoose}}} &default &\cellcolor[HTML]{ea9999}Failed &\cellcolor[HTML]{ea9999}Failed &\cellcolor[HTML]{ea9999}Failed &\cellcolor[HTML]{ea9999}Failed &\cellcolor[HTML]{ea9999}Failed &\cellcolor[HTML]{ea9999}Failed &\multirow{2}{*}{\textbf{31.55}} &\multirow{2}{*}{67.86} \\
&accurate &1094.18 &66738 &468.43 &48.78 &1562.62 &1142.97 & & \\
\bottomrule
\end{NiceTabularX}
\caption{Run-time statistics. For previous work, we include the tetrahedral meshing algorithm, either tetgen,  tetwild (default), or tetwild (accurate), the number of interior quadrature points, BBW and QHW optimization time, total run-times for BBW and QHW including tetrahedral meshing. Cells highlighted in red signify crashes, whereas cells highlighted in yellow are cases where the FEM pipeline produced a result, but either the result was unusable (\Gear{} mesh) or tetgen was unable to get within $10$\% of the $64$k interior quadrature points. For our work, we include timings with and without Neumann boundary conditions.}\label{tbl:qant}
\end{table*}

For a quantitative evaluation, Figure~\ref{fig:msloss} shows the energy during optimization for three different meshes, comparing directly optimizing a high-resolution $\mathcal X$ to using the multiscale strategy.
To make the plots comparable, we make sure that the final stage of the multiscale optimization has the same number of points as the high-resolution setup, and compute the energies using a finite difference estimator with $h$ equal to the one used for the high-resolution scenario.
To decrease the visual noise, we use a standard axis-aligned finite-difference estimator for computing the plots only.

As evidenced by this experiment, low-resolution solutions---i.e., the first stage of the the multiscale optimization---are quick to converge, but can result in a higher energy than the high-resolution solutions such as for the \Banana{} mesh.
Directly optimizing the high-resolution mesh results in slow convergence and overall higher error in all three scenarios.

These plots are instructive in other ways; for example, it is likely that optimizing for longer would further decrease the energy in all cases.
Similarly, it is likely that upsampling sooner in the first two stages would have been possible with little impact to the quality.
We found the current parameters to offer a good trade-off between simplicity of implementation, speed, and quality, and leave further experimentation to future work.

\subsection{Comparisons with Finite Elements}\label{ss:comp}

\paragraph*{Experimental Setup} We first attempt to tetrahedralize the boundary mesh with tetgen \cite{tetgen}, including samples on the control handles.
As QHW requires ellipsoidal handles, we first convert bones and points into ellipsoids with width equal to the width of our Lagrange constraint mollifier $\varepsilon$ from Equation~\ref{eq:lagrmollifier}.
If tetgen fails, we then attempt to tetrahedralize with FastTetWild~\cite{ftetwild}.
As FastTetWild cannot incorporate prescribed quadrature points into the mesh, we refine it with tetgen to make it usable for skinning.

Due to the many hyperparameters in tetrahedralization software, direct comparisons to our method are difficult.
The first important parameter is how well FastTetWild adheres to the boundary mesh.
As demonstrated by Figure~\ref{fig:tetfail}, the default parameters forego mesh quality for speed, resulting in low-resolution and occasionally incorrect solutions. 
FastTetWild can increase boundary adherence, at the expense of longer runtimes.
We include timings for both parameter settings, default and accurate, in Table~\ref{tbl:qant} and Figure~\ref{fig:timings}.

Similarly, tetgen offers a parameter to control the quality---and therefore density---of the tetrahedral mesh.
The default parameter used by \citet{BBW:2011} yields meshes of varying density, ranging from only $6244$ internal vertices for \textsc{Fish} to $150713$ for \textsc{Cow}.
Both differ from the total number of points used by our method in $3$D, $64000$.
To facilitate a fair comparison, we ran our experiments twice, once with default tetgen parameters and once attempting to match the number of internal vertices to our method.
Since the number of vertices is not exposed as a parameter in tetgen, we ran a binary search on the ``quality'' parameter and picked the result closest to $64000$.
We were unable to match the internal mesh density in three cases  (highlighted yellow in Table~\ref{tbl:qant}).
We include the results for the matched vertex count in Table~\ref{tbl:qant} and Figure~\ref{fig:timings}, and include the remaining results and raw data in the supplemental material.

\begin{figure}[hb!]
  \centering
  \includegraphics[width=\linewidth]{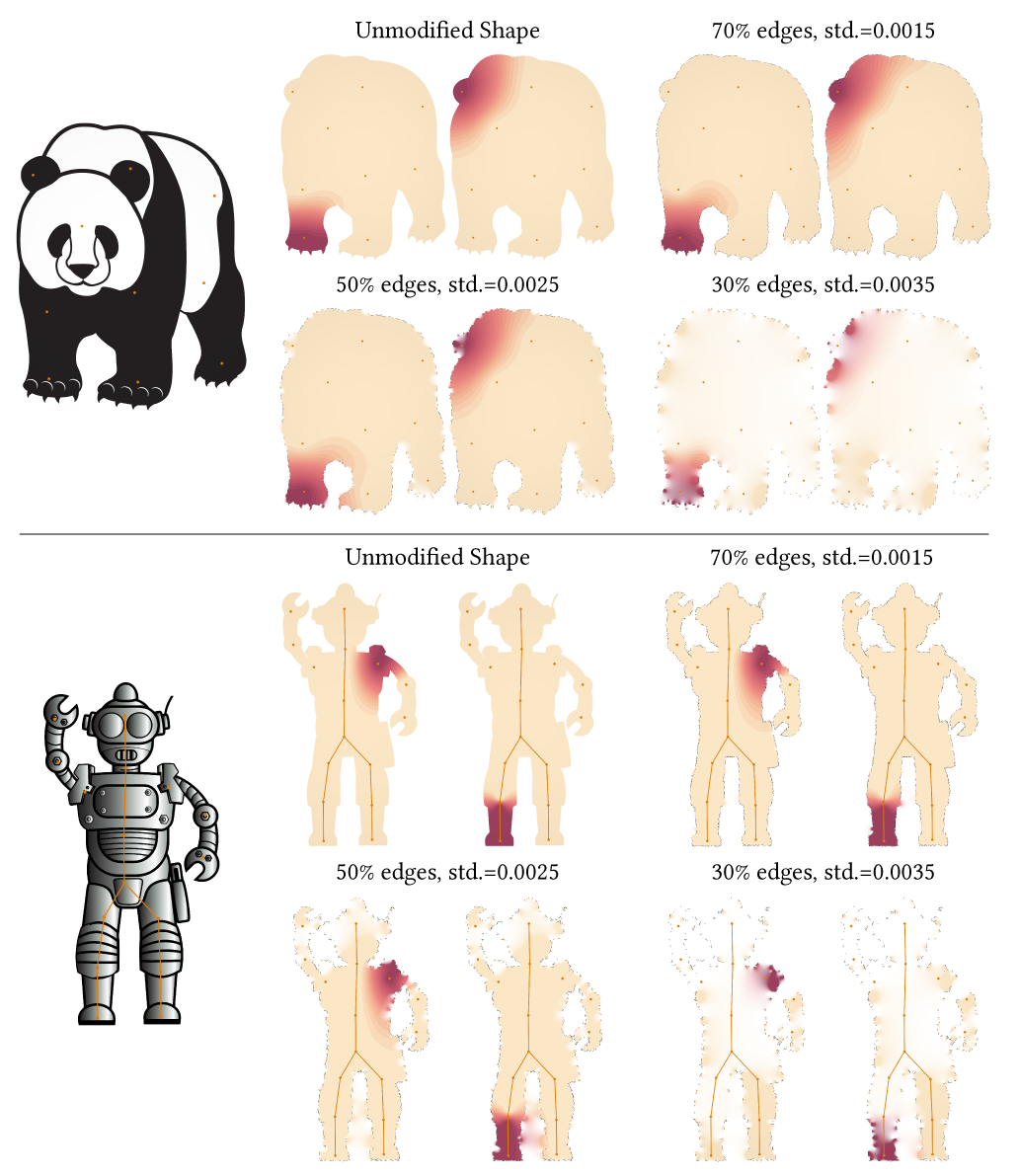}
  \caption{Our method is effective even on heavily degraded geometric data. This figure shows the results of our method on two shapes with progressively fewer boundary edges whose vertices have been randomly perturbed by adding progressively larger Gaussian noise.}
  \Description{Two different 2D meshes, each with different levels of boundary degredation: $70\%$ of edges, $50\%$ of edges, $30\%$ of edges.}\label{fig:soup}
\end{figure}

\begin{figure}[ht!]
  \centering
  \includegraphics[width=\linewidth]{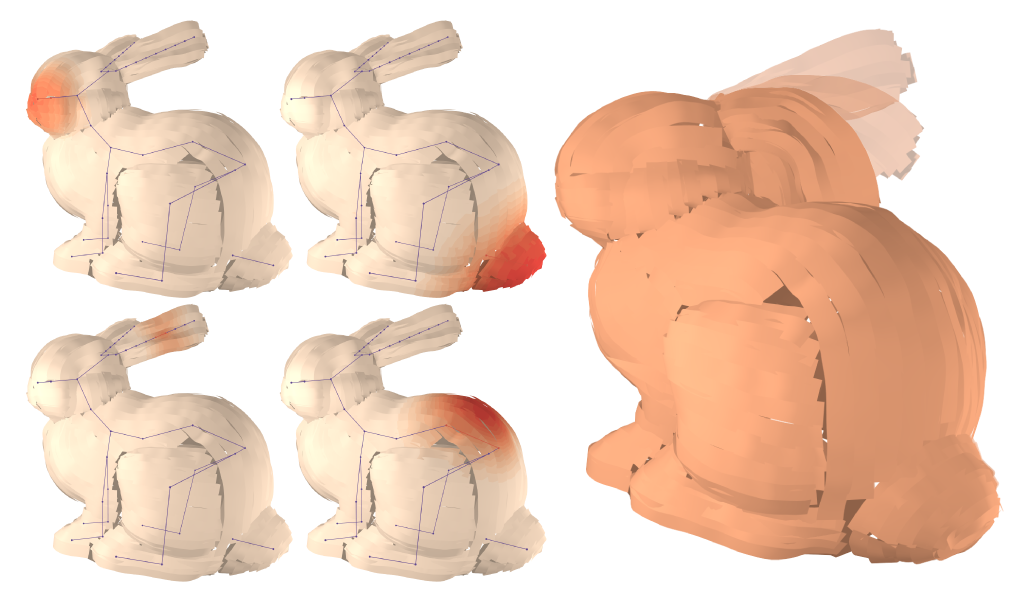}
  \caption{Our method allows us to animate non-watertight, self-intersecting shapes, such as these virtual reality ribbon drawings created in \textsc{SurfaceBrush} \cite{rosales2019SurfaceBrush}. Despite the quality of the geometry, our method produces smooth-looking weights and animations.}
  \Description{Visualization of skinning weights and animation on the bunny ribbon mesh.}\label{fig:bunny}
\end{figure}

\paragraph*{Findings} Five meshes, \Hand{}, \Penguin{}, \Fish{}, \Mushroom{}, and \Shiba{}, can be tetrahedralized with tetgen.
While our method is faster than BBW on these meshes, QHW is faster in comparison to our method.
However, once meshes have to be tetrahedralized with FastTetWild and refined with tetgen, robustness becomes a large issue.
Default FastTetWild parameters result in the BBW and QHW pipelines failing to produce results on three meshes (\textsc{Scorpion (Randomized)}, \textsc{Beaver}, and \textsc{Goosemoose}) and yield unusable weights on the \Gear{} mesh (Figure~\ref{fig:tetfail})
Accurate FastTetWild settings lead to tetrahedralization crashes on four meshes: \Scorpion{}, \textsc{Scorpion (Randomized)}, \textsc{Beaver}, and \Bunny{}.
In addition, QHW (but not BBW) crashes on the \textsc{Cow} mesh on the accurate FastTetWild setting.
\Piggybank{} is the \emph{only} FastTetWild mesh that succeeds under all experimental settings.
In the remaining cases, our method is either comparable to, or faster than previous work.
Figures~\ref{fig:teaser}, \ref{fig:tetfail}, \ref{fig:companim}, \ref{fig:soup}, \ref{fig:bunny}, and \ref{fig:open} demonstrate these key findings qualitatively.

Our main contribution, i.e., effectiveness on imperfect data, is exemplified by the extreme example of triangle soups in Figures~\ref{fig:teaser}~and~\ref{fig:soup}, as well as others in Figure~\ref{fig:open}.
Despite severe mesh deterioration, our method still produces meaningful solutions resulting in smooth looking animations.
Figure~\ref{fig:soup} demonstrates the graceful degradation of our method inherited from the generalized winding number; even with $70$\% of the polygon edges removed and the remaining ones randomly displaced, our method succeeds.
In Figure~\ref{fig:teaser}, we use our weights to animate \textsc{Scorpion (Randomized)}, obtained by deteriorating \Scorpion{}; FEM-based algorithms are unable to produce a solution for such a degraded mesh. %

While triangle soups are an extreme example, algorithms relying on tetrahedral meshing fail in more common scenarios.
For example, they fail on at least one experimental setting on \Scorpion{} (Figure~\ref{fig:companim}), \textsc{Goosemoose} (Figure~\ref{fig:goosemoose}), and \textsc{Cow} (Figure~\ref{fig:cow}) meshes, despite no obvious problems with the meshes.
BBW and QHW results on the \Gear{} mesh either contain artifacts that render the result unusable or require a $1.66$ hour meshing time.

Two practically-relevant classes of shapes our method handles out-of-the box are $3$D scans from computer vision pipelines (Figure~\ref{fig:teaser}) and virtual reality ribbon drawings created by e.g.\ \textsc{SurfaceBrush} \cite{rosales2019SurfaceBrush} (Figures~\ref{fig:teaser}~and~\ref{fig:bunny}).
Both of these modalities produce non-watertight, self-intersecting meshes, which necessitate slow and brittle tetrahedralization software.
The $3$D scan of the \textsc{Beaver} mesh (Figure~\ref{fig:teaser}) fails to tetrahedralize, regardless of settings, \Bunny{} (Figure~\ref{fig:bunny}) either fails to tetrahedralize or results in a $7.25$ minute runtime for QHW and a $18.32$ minute runtime for BBW (in comparison to our method's $71.74$ seconds).
While \Piggybank{} (Figure~\ref{fig:teaser}) succeeds with all FEM-based methods, our method is nonetheless faster than the best QHW result on that mesh.

Similar to past work and most applications of BBW, we found that solving the full BBW quadratic program was prohibitively slow on realistic examples and consequently dropped the partition-of-unity constraint, which was enforced \emph{a posteriori} by dividing the weights by their sum pointwise.
This simplified problem was nonetheless often orders of magnitude slower than our method.

\begin{figure}[h]
  \centering
  \includegraphics[width=\linewidth]{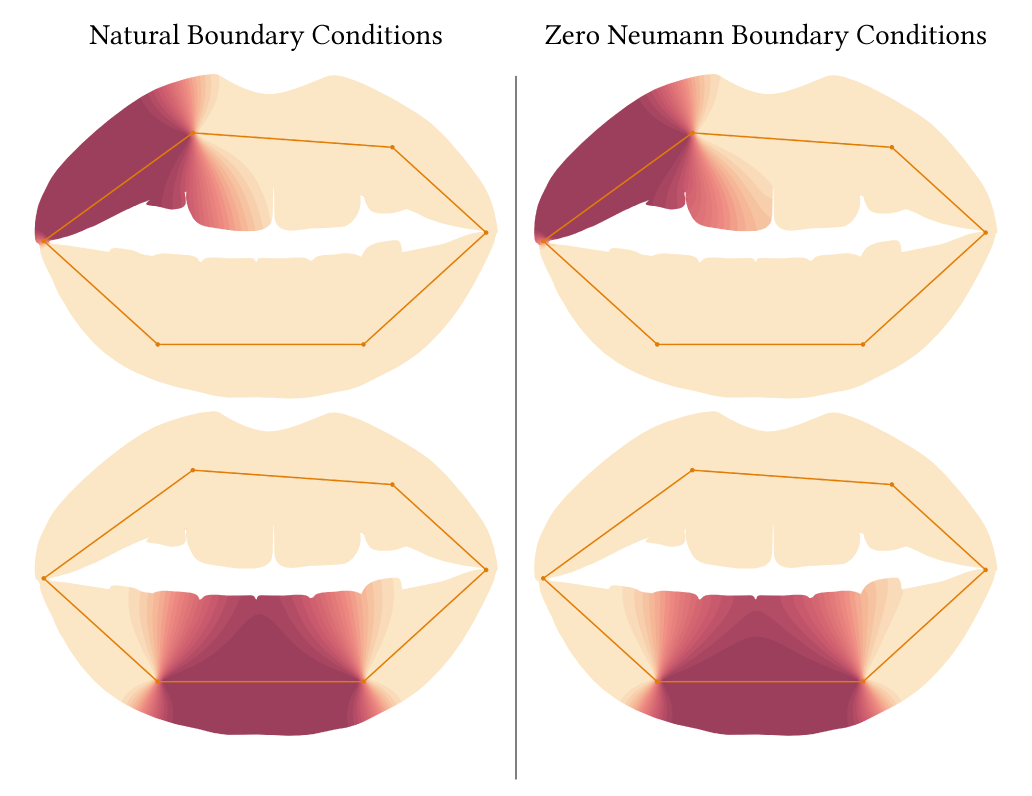}
  \caption{We demonstrate the effect of the optional zero Neumann boundary condition constraints; on the right, the isolines of the level-sets are indeed orthogonal to the boundary of the shape.}
  \Description{2D mesh with and without zero Neumann conditions.}\label{fig:noumann}
\end{figure}

\subsection{Boundary Conditions}\label{ss:bcres}

Next, we demonstrate our method's flexibility to boundary conditions.
Figures~\ref{fig:bananeuman}~and~\ref{fig:neumanbc} visualize the difference between natural boundary conditions and the optional Neumann conditions.
Figure~\ref{fig:bananeuman} provides a concrete example where Neumann conditions might be desirable, as they make the weights symmetric through narrow passages.
We include all of the $3$D results with and without Neumannn boundary conditions as supplemental material.
The differences between the boundary conditions are noticeable albeit subtle, but the natural boundary conditions are significantly faster than the Neumann conditions (Table~\ref{tbl:qant}).
We leave enabling or disabling the zero Neumann conditions as an optional choice for the user.

\begin{figure}[ht!]
  \centering
  \includegraphics[width=\linewidth]{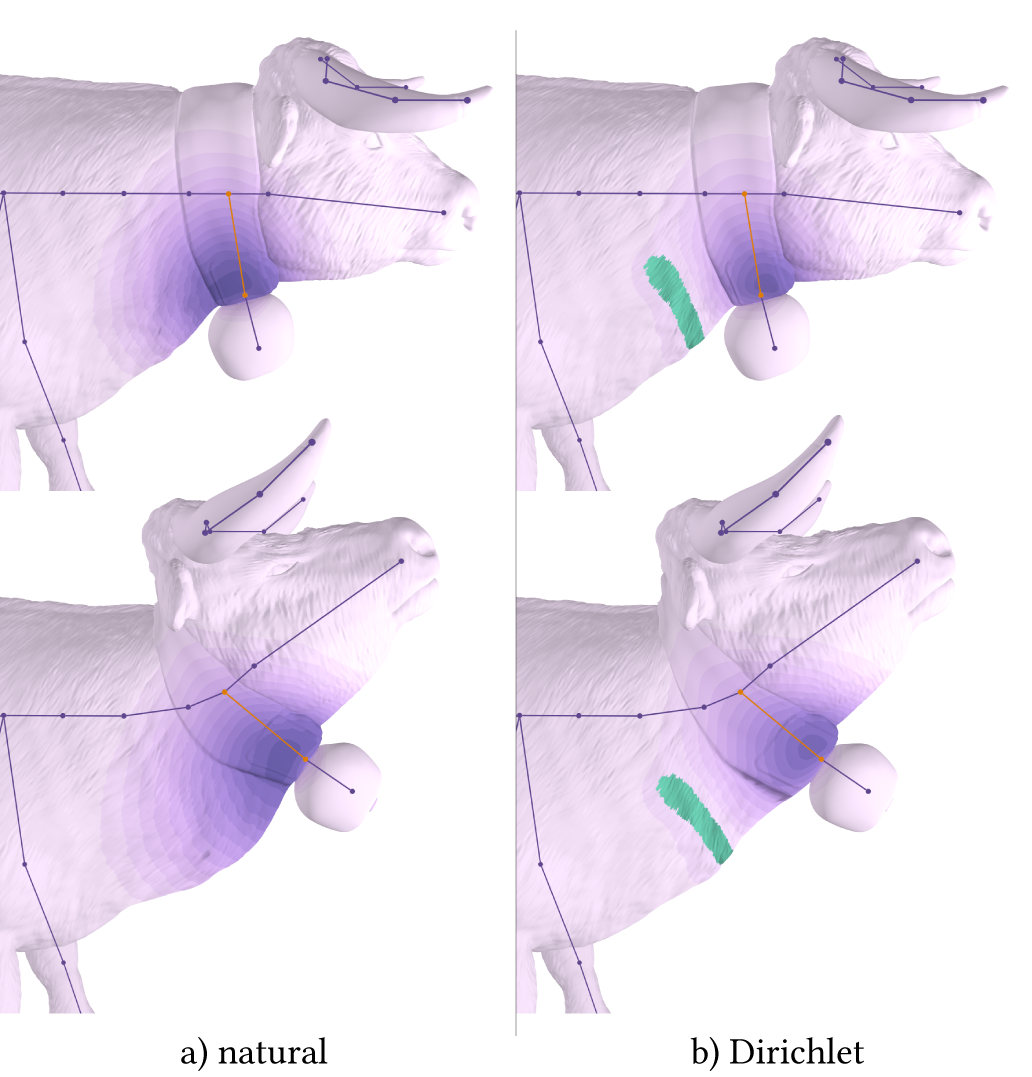}
  \caption{Optimizing bounded biharmonic weights with natural boundary conditions offers little control in the final look of the animation (a). By pausing the optimization after $500$ iterations and painting Dirichlet conditions onto a small region (pictured in green), the user can, for example, attenuate a handle's influence over said region (b).}
  \Description{3D mesh with and without hand painted Dirichlet conditions on a subset of the mesh.}\label{fig:cow}
\end{figure}

\begin{figure}[hb!]
  \centering
  \includegraphics[width=\linewidth]{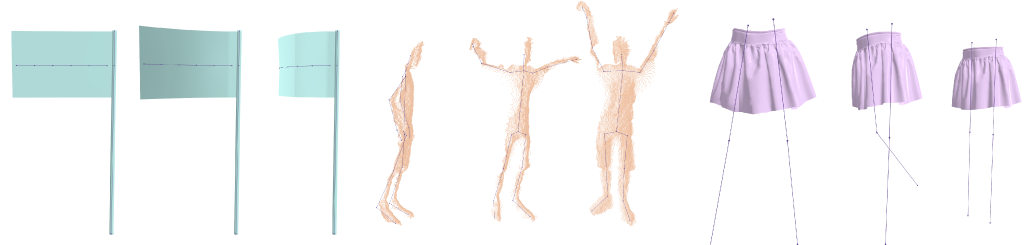}
  \caption{Our method succeeds in computing weights in extreme stress test scenarios including a flag, a partial point cloud of a human from a smartphone depth sensor, and a skirt mesh with multiple disconnected patches.}
  \Description{Animations of a flag, a point cloud, and a skirt.}\label{fig:open}
\end{figure}

More interesting are the Dirichlet conditions, which open new connections between bounded biharmonic weights and weight painting, the predominant user interaction for skinning weight design.
As demonstrated in Figure~\ref{fig:goosemoose}, painted weights can allow users to connect disconnected components to a given handle.
Figure~\ref{fig:cow} demonstrates how painting can also allow for artistic control.
This boundary condition is a crucial missing puzzle piece for optimization-based automatic skinning weight design.

\subsection{Discussion}\label{ss:disc}

Our results show conclusively that geometric fields are useful for the computation of bounded biharmonic weights. 
While QHW still produces results somewhat quicker than our approach, our method is more robust and faster on meshes requiring robust tetrahedralization.
Our formulation also introduces useful new controls to the bounded biharmonic weights in the form of boundary conditions.
Dirichlet conditions are especially important, as they allow for manual user intervention to produce a specific result.

\paragraph*{Hyperparameters.} Our method includes some parameters, namely the resolution of the underlying representation, sizes of boundary condition mollifiers, and optimizer parameters.
While these parameters are different than the ones used by BBW and QHW (which require parameters to the tet meshing tool and optimization software), we have a similar number of parameters to past work, and our default parameters are effective across all examples.

Controlling the exact resolution of the tetrahedral meshes in our comparisons proved difficult in our experiments and required a binary search over a ``quality" parameter.
As the different experimental settings demonstrate, we were unable to find a consistent setting of the meshing tools' parameters that make previous methods produce usable results on all meshes.

\paragraph*{Limitations and Future Work}~\label{ss:limit} %
Our boundary condition mollifiers are small but finite, and they are not geometry aware.
This simple construction could result in unintuitive results if, for example, there were a bone placed extremely close to the boundary the \textsc{Gear} mesh, as this situation would cause bleeding.
We did not observe this situation in our examples, since the width of the boundary conditions decreases as the underlying resolution increases (see Appendix~\ref{s:impl}).
This boundary case could be resolved by tracing rays to the closest point on the bone to check visibility, automatically determining the size of the boundary conditions based on the boundary geometry, or by adding a way to detect this case and informing the user.

\begin{figure}[hb!]
  \centering
  \includegraphics[width=\linewidth]{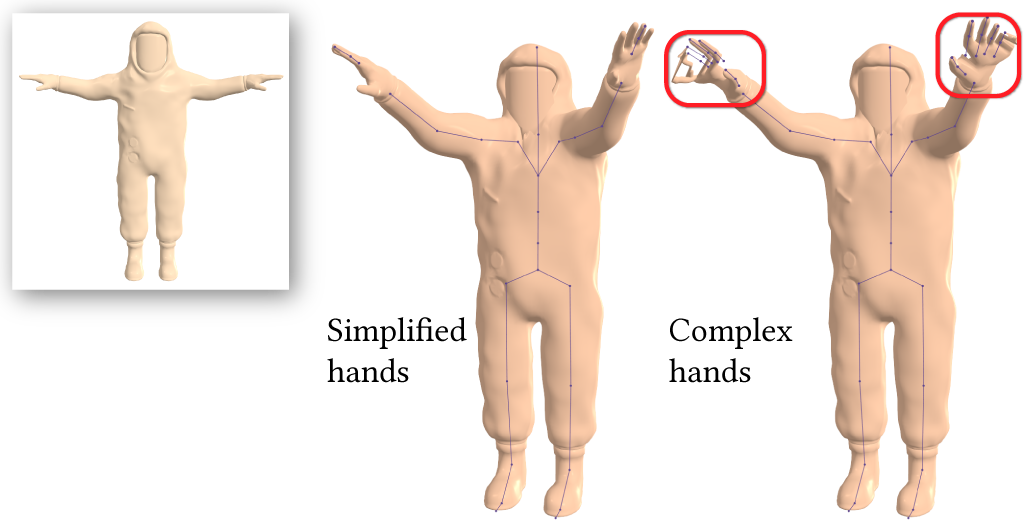}
  \caption{The body of the person depicted in the \textsc{Hazmat} mesh is large compared to their fingers; the undeformed mesh is shown on the left. Despite disconnected and self-intersecting geometry, our method works as expected if we care to animate the body alone (center). If we need to animate each finger separately using a high-resolution hand rig (right), our method still produces a result in under $30$ seconds, but the animations include visible artifacts. See discussion in Section~\ref{ss:disc}.}
  \Description{Humanoid mesh deformation, once with a single bone joint in the hands and once with a fully articulated hand. The fully articulated hand has noticeable artifacts.}\label{fig:hazmat}
\end{figure}

\begin{figure}[ht!]
  \centering
  \includegraphics[width=\linewidth]{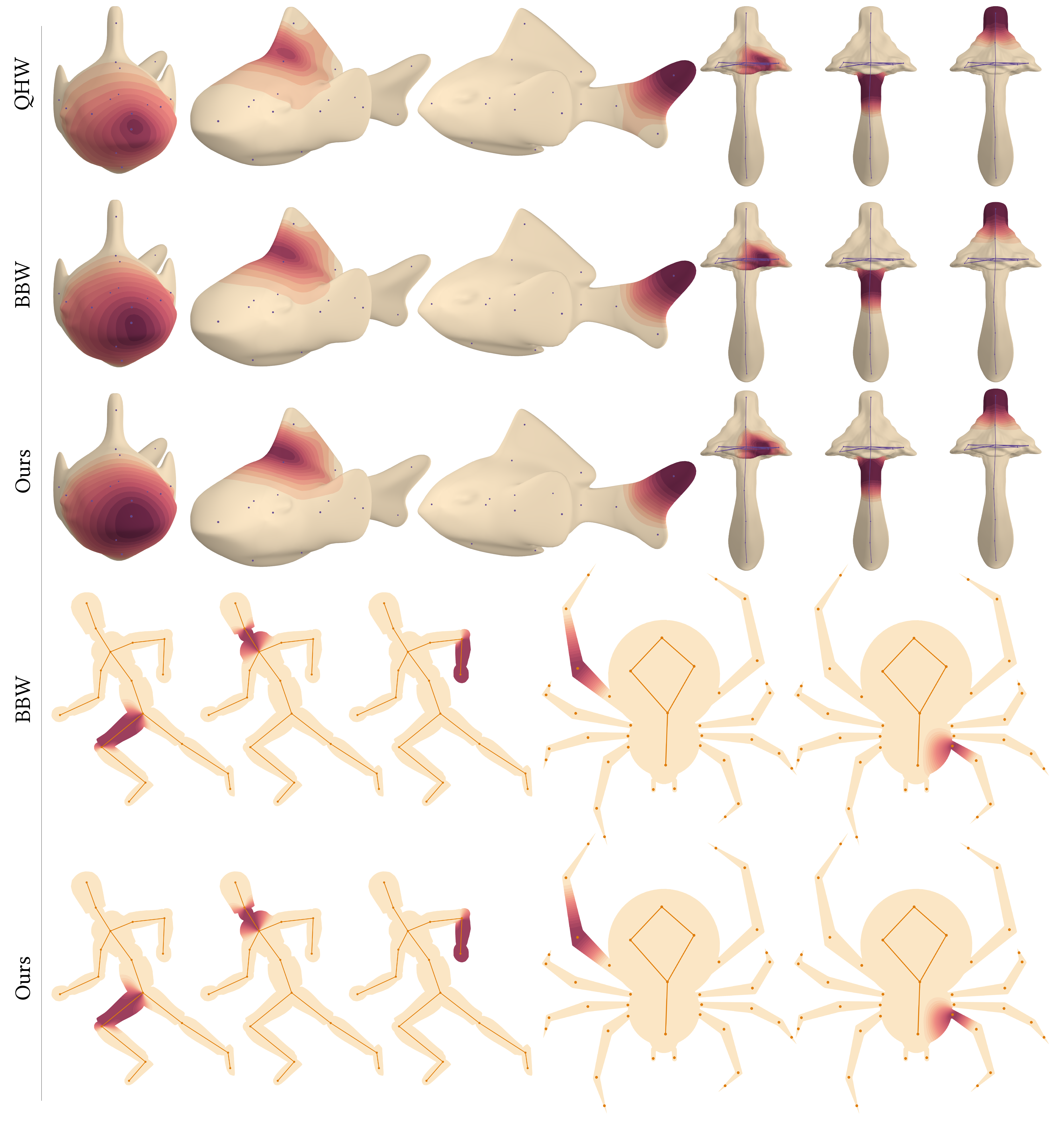}
  \caption{Comparison of our results to quasi-harmonic weights QHW \cite{wang2021fast} and bounded biharmonic weights (BBW)~\cite{BBW:2011}.  QHW and BBW rely on tetrahedralization in $3$D or triangulation in $2$D, while our method requires only the boundary triangle mesh or polygon. All methods produce similar-looking weights as they are solving the same problem. Some minor differences are possible due to differences in discretization and enforcement of boundary conditions.}
  \Description{Qualitative comparison with previous methods on 2D and 3D meshes.}\label{fig:comparison}
\end{figure}

Similarly, there can be situations where the distance between bones is so small that the Lagrange property is broken \emph{if} our mollifiers and $\sigma$ are large-enough.
Note, however, that even in the continuous formulation \eqref{eq:bbw}, it is possible for the Lagrange property to be broken, if bones touch or intersect.
Furthermore, when animating filigree-like features that are small compared to the rest of the mesh, it can happen $\sigma$ is too large to meaningfully optimize the biharmonic energy in those regions.
Compounding this effect, small filigree regions can be poorly sampled at lower resolutions.
Such a case is demonstrated in Figure~\ref{fig:hazmat}; the individual bones of the human's fingers are too small and too close to each other compared to the rest of the human to be meaningfully represented by the point cloud at the lowest resolutions.
We anticipate that the solution to this will be importance sampling such that larger percentages of $\mathcal{X}$ and of the optimization samples are focused on narrow regions.`

Even though our method is reasonably fast, it involves ray-tracing hardware that is not yet ubiquitous; the majority of the runtime  is spent tracing rays.
While this limitation will become less of an issue as the availability of ray-tracing hardware increases in $3$D workstations, it is worth finding ways of amortizing this cost.

Even though our method is able to handle extreme scenarios, our random sampling relies on the generalized winding number for inside-outside testing and therefore inherits the preconditions of the generalized winding number, including a consistent orientation of normals.
However, our method is mostly orthogonal to the specific inside-outside testing algorithm---modifying it to work with a different robust inside-outside algorithm is likely possible, albeit out of the scope of the present works.

While we found our multiscale optimization strategy to be effective, we believe that the runtime could be further improved by finding a way to precondition geometric fields.
Methods such as QHW \cite{wang2021fast}, while using the Adam optimizer, rely on preconditioning for acceleration.

Lastly, our method is specifically tuned for bounded biharmonic weights, but one could imagine a general framework for solving variational problems using geometric fields.
Many problems in geometry processing can be formulated in terms of a variational principle, including, e.g., mesh parameterization and deformation, various formulations of developable surface approximation, geometric flows, etc.
In the specific context of the Dirichlet energy, future work might want to connect to the work of \citet{Sawhney:2020:MCG}, where our representation could be used as a direct replacement, or---when convergence guarantees are needed---as a control variate.

\section{Conclusion}
We propose a mesh-free approach to automatic computation of skinning weights. This is enabled by our geometric field representation of skinning weights, through which we may optimize the biharmonic energy without the need for finite elements or constraints. Our method enables skinning weights computation on previously challenging or even impossible shapes, such as those with open surfaces or triangle soups, while achieving quality comparable to prior state-of-the-art approaches requiring volumetric meshes. 

\begin{acks}
We thank Yu Wang for helping us compare with quasi-harmonic weights, and to Enrique Rosales for the permission to use the \Piggybank{} and \Bunny{} meshes.
Thank you to Ahmed Mahmoud, David Palmer, and Silvia Sell\'an for help with proof reading.

The MIT Geometric Data Processing Group acknowledges the generous support of Army Research Office grants W911NF2010168 and W911NF2110293, of Air Force Office of Scientific Research award FA9550-19-1-031, of National Science Foundation grant CHS-1955697, from the CSAIL Systems that Learn program, from the MIT–IBM Watson AI Laboratory, from the Toyota–CSAIL Joint Research Center, from a gift from Adobe Systems, and from a Google Research Scholar award.

The MIT Scene Representation Group acknowledges the generous support of the National Science Foundation under Grant No. 2211259, of the Singapore DSTA under DST00OECI20300823 (New Representations for Vision), of the Intelligence Advanced Research Projects Activity (IARPA) via Department of Interior/ Interior Business Center (DOI/IBC) under 140D0423C0075, of the Amazon Science Hub, and of IBM.

Oded Stein acknowledges the generous support of the SNF Early Postdoc Mobility fellowship and of the National Science Foundation (award \#2335493).
\end{acks}

\bibliographystyle{ACM-Reference-Format}
\bibliography{bibliography}

@String{Computing = "Computing" }

@String{Computer = "{IEEE} Computer" }

@String{Springer = "Springer-Verlag" }

@Article{Chen2024ahard,
AUTHOR = {Chen, Simin and Liu, Zhixiang and Zhang, Wenbo and Yang, Jinkun},
TITLE = {A Hard-Constraint Wide-Body Physics-Informed Neural Network Model for Solving Multiple Cases in Forward Problems for Partial Differential Equations},
JOURNAL = {Applied Sciences},
VOLUME = {14},
YEAR = {2024},
NUMBER = {1},
ARTICLE-NUMBER = {189}
}

@inproceedings{anonymous2024scaling,
  title={Scaling physics-informed hard constraints with mixture-of-experts},
  author={Nithin Chalapathi and Yiheng Du and Aditi S. Krishnapriyan},
  booktitle={The Twelfth International Conference on Learning Representations},
  year={2024},
  url={https://openreview.net/forum?id=u3dX2CEIZb}
}

@misc{djeumou2022neural,
      title={Neural Networks with Physics-Informed Architectures and Constraints for Dynamical Systems Modeling}, 
      author={Franck Djeumou and Cyrus Neary and Eric Goubault and Sylvie Putot and Ufuk Topcu},
      year={2022},
      eprint={2109.06407},
      archivePrefix={arXiv},
      primaryClass={cs.LG}
}

@article{Mohan2023,
  title = {Embedding hard physical constraints in neural network coarse-graining of three-dimensional turbulence},
  author = {Mohan, Arvind T. and Lubbers, Nicholas and Chertkov, Misha and Livescu, Daniel},
  journal = {Phys. Rev. Fluids},
  volume = {8},
  issue = {1},
  numpages = {17},
  year = {2023}
}

@misc{rodrigueztorrado2021physicsinformed,
      title={Physics-informed attention-based neural network for solving non-linear partial differential equations}, 
      author={Ruben Rodriguez-Torrado and Pablo Ruiz and Luis Cueto-Felgueroso and Michael Cerny Green and Tyler Friesen and Sebastien Matringe and Julian Togelius},
      year={2021},
      eprint={2105.07898},
      archivePrefix={arXiv},
      primaryClass={cs.LG}
}

@inproceedings{Liu2022aunified,
 author = {Liu, Songming and Zhongkai, Hao and Ying, Chengyang and Su, Hang and Zhu, Jun and Cheng, Ze},
 booktitle = {Advances in Neural Information Processing Systems},
 pages = {20287--20299},
 title = {A Unified Hard-Constraint Framework for Solving Geometrically Complex PDEs},
 volume = {35},
 year = {2022}
}

@article{jeruzalski2020nilbs,
  title   = {{NiLBS}: Neural inverse linear blend skinning},
  author  = {Jeruzalski, Timothy and Levin, David IW and Jacobson, Alec and Lalonde, Paul and Norouzi, Mohammad and Tagliasacchi, Andrea},
  journal = {arXiv:2004.05980},
  year    = {2020}
}

@article{kingma2014adam,
  author    = {Kingma, Diederik P and Ba, Jimmy},
  journal   = {arXiv:1412.6980},
  title     = {Adam: A method for stochastic optimization},
  year      = 2014
}

@incollection{pytorch,
  title     = {PyTorch: An Imperative Style, High-Performance Deep Learning Library},
  author    = {Paszke, Adam and Gross, Sam and Massa, Francisco and Lerer, Adam and Bradbury, James and Chanan, Gregory and Killeen, Trevor and Lin, Zeming and Gimelshein, Natalia and Antiga, Luca and Desmaison, Alban and Kopf, Andreas and Yang, Edward and DeVito, Zachary and Raison, Martin and Tejani, Alykhan and Chilamkurthy, Sasank and Steiner, Benoit and Fang, Lu and Bai, Junjie and Chintala, Soumith},
  booktitle = {Advances in Neural Information Processing Systems 32},
  pages     = {8024--8035},
  year      = {2019},
  publisher = {Curran Associates, Inc.},
  url       = {http://papers.neurips.cc/paper/9015-pytorch-an-imperative-style-high-performance-deep-learning-library.pdf}
}

@article{Sukumar2022,
  title   = {Exact imposition of boundary conditions with distance functions in physics-informed deep neural networks},
  journal = {Computer Methods in Applied Mechanics and Engineering},
  volume  = {389},
  pages   = {114333},
  year    = {2022},
  author  = {N. Sukumar and Ankit Srivastava}
}

@inproceedings{xie2022neural,
  title        = {Neural fields in visual computing and beyond},
  author       = {Xie, Yiheng and Takikawa, Towaki and Saito, Shunsuke and Litany, Or and Yan, Shiqin and Khan, Numair and Tombari, Federico and Tompkin, James and Sitzmann, Vincent and Sridhar, Srinath},
  booktitle    = {Computer Graphics Forum},
  volume       = {41},
  number       = {2},
  pages        = {641--676},
  year         = {2022},
  organization = {Wiley Online Library}
}

@article{Dodik:VBC:2023,
    title = {Variational Barycentric Coordinates},
    doi = {https://doi.org/10.1145/3618403},
    author = {Ana Dodik and Oded Stein and Vincent Sitzmann and Justin Solomon},
    year = {2023},
    journal = {ACM Transactions on Graphics}
}

@article{BBW:2011,
author = {Alec Jacobson and Ilya Baran and Jovan Popovi{\'{c}} and Olga Sorkine},
title = {Bounded Biharmonic Weights for Real-Time Deformation},
journal = {ACM Transactions on Graphics (proceedings of ACM SIGGRAPH)},
volume = {30},
number = {4},
year = {2011},
pages = {78:1--78:8},
}

@article{Jacobson:MONO:2012,
author = {Alec Jacobson and Tino Weinkauf and Olga Sorkine},
title = {Smooth Shape-Aware Functions with Controlled Extrema},
journal = {Computer Graphics Forum (proceedings of EUROGRAPHICS/ACM SIGGRAPH Symposium on Geometry Processing)},
volume = {31},
pages = {1577--1586},
number = {5},
year = {2012},
}

@article{barill2018fast,
author = {Barill, Gavin and Dickson, Neil G. and Schmidt, Ryan and Levin, David I. W. and Jacobson, Alec},
title = {Fast winding numbers for soups and clouds},
year = {2018},
issue_date = {August 2018},
publisher = {Association for Computing Machinery},
address = {New York, NY, USA},
volume = {37},
number = {4},
issn = {0730-0301},
url = {https://doi.org/10.1145/3197517.3201337},
doi = {10.1145/3197517.3201337},
journal = {ACM Trans. Graph.},
month = jul,
articleno = {43},
numpages = {12}
}

@ARTICLE{Kavan08DQS,
  AUTHOR       = {Ladislav Kavan and Steven Collins and Jiri Zara and Carol O'Sullivan},
  TITLE        = {Geometric Skinning with Approximate Dual Quaternion Blending},
  JOURNAL      = {ACM Trans. Graph.},
  VOLUME       = {27},
  NUMBER       = {4},
  YEAR         = {2008},
  PUBLISHER    = {ACM Press},
  PAGES        = {105},
  ADDRESS      = {New York, NY, USA},
}

@article{tetwild,
author = {Hu, Yixin and Zhou, Qingnan and Gao, Xifeng and Jacobson, Alec and Zorin, Denis and Panozzo, Daniele},
title = {Tetrahedral Meshing in the Wild},
year = {2018},
issue_date = {August 2018},
publisher = {Association for Computing Machinery},
address = {New York, NY, USA},
volume = {37},
number = {4},
issn = {0730-0301},
url = {https://doi.org/10.1145/3197517.3201353},
doi = {10.1145/3197517.3201353},
journal = {ACM Trans. Graph.},
month = {jul},
articleno = {60},
numpages = {14}
}

@article{ftetwild,
author = {Hu, Yixin and Schneider, Teseo and Wang, Bolun and Zorin, Denis and Panozzo, Daniele},
title = {Fast Tetrahedral Meshing in the Wild},
year = {2020},
issue_date = {July 2020},
publisher = {Association for Computing Machinery},
address = {New York, NY, USA},
volume = {39},
number = {4},
issn = {0730-0301},
url = {https://doi.org/10.1145/3386569.3392385},
doi = {10.1145/3386569.3392385},
journal = {ACM Trans. Graph.},
month = jul,
articleno = {117},
numpages = {18}
}

@article{Diazzi:2023,
author = {Diazzi, Lorenzo and Panozzo, Daniele and Vaxman, Amir and Attene, Marco},
title = {Constrained Delaunay Tetrahedrization: A Robust and Practical Approach},
year = {2023},
issue_date = {December 2023},
publisher = {Association for Computing Machinery},
address = {New York, NY, USA},
volume = {42},
number = {6},
issn = {0730-0301},
url = {https://doi.org/10.1145/3618352},
doi = {10.1145/3618352},
journal = {ACM Trans. Graph.},
month = {dec},
articleno = {181},
numpages = {15}
}

@article{Nadaraya:1964,
author = {Nadaraya, E. A.},
title = {On Estimating Regression},
journal = {Theory of Probability \& Its Applications},
volume = {9},
number = {1},
pages = {141-142},
year = {1964},
doi = {10.1137/1109020},
URL = {https://doi.org/10.1137/1109020},
eprint = {https://doi.org/10.1137/1109020}
}

@article{Watson:1964,
 ISSN = {0581572X},
 URL = {http://www.jstor.org/stable/25049340},
 author = {Geoffrey S. Watson},
 journal = {Sankhyā: The Indian Journal of Statistics, Series A (1961-2002)},
 number = {4},
 pages = {359--372},
 publisher = {Springer},
 title = {Smooth Regression Analysis},
 urldate = {2024-01-19},
 volume = {26},
 year = {1964}
}

@article{CoifmanLafon:2006,
title = {Diffusion maps},
journal = {Applied and Computational Harmonic Analysis},
volume = {21},
number = {1},
pages = {5-30},
year = {2006},
note = {Special Issue: Diffusion Maps and Wavelets},
issn = {1063-5203},
doi = {https://doi.org/10.1016/j.acha.2006.04.006},
url = {https://www.sciencedirect.com/science/article/pii/S1063520306000546},
author = {Ronald R. Coifman and Stéphane Lafon}
}

@book{PBRT3e,
    title = {Physically Based Rendering: From Theory to Implementation (3rd ed.)},
    author = {Matt Pharr and Wenzel Jakob and Greg Humphreys},
    isbn = {9780128006450},
    pages = {1266},
    year = {2016},
    month = {oct},
    edition = {3rd},
    publisher = {Morgan Kaufmann Publishers Inc.},
    address = {San Francisco, CA, USA}
}

@article{Parker10OptiX,
 author = {Steven G. Parker and James Bigler and Andreas Dietrich and Heiko Friedrich and Jared Hoberock and David Luebke and  David McAllister and Morgan McGuire and Keith Morley and Austin Robison and Martin Stich},
 title = {OptiX: A General Purpose Ray Tracing Engine},
 year = {2010},
 month = {August},
 journal = {ACM Transactions on Graphics}
 }

@article{tetgen,
author = {Si, Hang},
title = {TetGen, a Delaunay-Based Quality Tetrahedral Mesh Generator},
year = {2015},
issue_date = {January 2015},
publisher = {Association for Computing Machinery},
address = {New York, NY, USA},
volume = {41},
number = {2},
issn = {0098-3500},
url = {https://doi.org/10.1145/2629697},
doi = {10.1145/2629697},
journal = {ACM Trans. Math. Softw.},
month = {feb},
articleno = {11},
numpages = {36}
}

@article{Pauly:2003,
author = {Pauly, Mark and Keiser, Richard and Kobbelt, Leif P. and Gross, Markus},
title = {Shape modeling with point-sampled geometry},
year = {2003},
issue_date = {July 2003},
publisher = {Association for Computing Machinery},
address = {New York, NY, USA},
volume = {22},
number = {3},
issn = {0730-0301},
url = {https://doi.org/10.1145/882262.882319},
doi = {10.1145/882262.882319},
journal = {ACM Trans. Graph.},
month = {jul},
pages = {641–650},
numpages = {10}
}

@Article{Kerbl2023,
      author       = {Kerbl, Bernhard and Kopanas, Georgios and Leimk{\"u}hler, Thomas and Drettakis, George},
      title        = {3D Gaussian Splatting for Real-Time Radiance Field Rendering},
      journal      = {ACM Transactions on Graphics},
      number       = {4},
      volume       = {42},
      year         = {2023}
}

@inproceedings{Alexa:2004,
author = {Alexa, Marc and Gross, Markus and Pauly, Mark and Pfister, Hanspeter and Stamminger, Marc and Zwicker, Matthias},
title = {Point-based computer graphics},
year = {2004},
isbn = {9781450378017},
publisher = {Association for Computing Machinery},
address = {New York, NY, USA},
url = {https://doi.org/10.1145/1103900.1103907},
doi = {10.1145/1103900.1103907},
booktitle = {ACM SIGGRAPH 2004 Course Notes},
pages = {7–es},
location = {Los Angeles, CA},
series = {SIGGRAPH '04}
}

@inproceedings{halen2021global,
  title={Global illumination based on surfels},
  author={Halen, Henrik and Hayward, K},
  booktitle={Proc. ACM SIGGRAPH Symp. Interactive 3D Graph. Games},
  pages={1399--1405},
  year={2021}
}

@article{Majercik2019Irradiance,
  author =       {Zander Majercik and Jean-Philippe Guertin and Derek Nowrouzezahrai and Morgan McGuire}, 
  title =        {Dynamic Diffuse Global Illumination with Ray-Traced Irradiance Fields},
  year =         2019,
  month =        {June},
  day =          5,
  journal =      {Journal of Computer Graphics Techniques (JCGT)},
  volume =       8,
  number =       2,
  pages =        {1--30},
  url =          {http://jcgt.org/published/0008/02/01/},
  issn =         {2331-7418}
}

@inproceedings{li2023neuralangelo,
  title={Neuralangelo: High-Fidelity Neural Surface Reconstruction},
  author={Li, Zhaoshuo and M\"uller, Thomas and Evans, Alex and Taylor, Russell H and Unberath, Mathias and Liu, Ming-Yu and Lin, Chen-Hsuan},
  booktitle={IEEE Conference on Computer Vision and Pattern Recognition ({CVPR})},
  year={2023}
}

@misc{chetan2023accurate,
      title={Accurate Differential Operators for Hybrid Neural Fields}, 
      author={Aditya Chetan and Guandao Yang and Zichen Wang and Steve Marschner and Bharath Hariharan},
      year={2023},
      eprint={2312.05984},
      archivePrefix={arXiv},
      primaryClass={cs.CV}
}

@article{Jacobson2013Winding,
author = {Alec Jacobson and Ladislav Kavan and Olga Sorkine},
title = {Robust Inside-Outside Segmentation using Generalized Winding Numbers},
journal = {ACM Trans. Graph.},
volume = {32},
number = {4},
year = {2013},
}

@article{james2005skinning,
  title={Skinning mesh animations},
  author={James, Doug L and Twigg, Christopher D},
  journal={ACM Transactions on Graphics (TOG)},
  volume={24},
  number={3},
  pages={399--407},
  year={2005},
  publisher={ACM New York, NY, USA}
}

@inproceedings{kavan2010fast,
  title={Fast and efficient skinning of animated meshes},
  author={Kavan, Ladislav and Sloan, Peter-Pike and O'Sullivan, Carol},
  booktitle={Computer Graphics Forum},
  volume={29},
  number={2},
  pages={327--336},
  year={2010},
  organization={Wiley Online Library}
}

@article{le2012smooth,
  title={Smooth skinning decomposition with rigid bones},
  author={Le, Binh Huy and Deng, Zhigang},
  journal={ACM Transactions on Graphics (TOG)},
  volume={31},
  number={6},
  pages={1--10},
  year={2012},
  publisher={ACM New York, NY, USA}
}

@article{le2014robust,
  title={Robust and accurate skeletal rigging from mesh sequences},
  author={Le, Binh Huy and Deng, Zhigang},
  journal={ACM Transactions on Graphics (TOG)},
  volume={33},
  number={4},
  pages={1--10},
  year={2014},
  publisher={ACM New York, NY, USA}
}

@article{wampler2016fast,
  title={Fast and reliable example-based mesh {IK} for stylized deformations},
  author={Wampler, Kevin},
  journal={ACM Transactions on Graphics (TOG)},
  volume={35},
  number={6},
  pages={1--12},
  year={2016},
  publisher={ACM New York, NY, USA}
}

@article{joshi2007harmonic,
  title={Harmonic coordinates for character articulation},
  author={Joshi, Pushkar and Meyer, Mark and DeRose, Tony and Green, Brian and Sanocki, Tom},
  journal={ACM Transactions on Graphics (TOG)},
  volume={26},
  number={3},
  pages={71-1--71-9},
  year={2007},
  publisher={ACM New York, NY, USA}
}

@article{baran2007automatic,
  title={Automatic rigging and animation of 3d characters},
  author={Baran, Ilya and Popovi{\'c}, Jovan},
  journal={ACM Transactions on graphics (TOG)},
  volume={26},
  number={3},
  pages={72-1--72-8},
  year={2007},
  publisher={ACM New York, NY, USA}
}

@article{botsch2004intuitive,
  title={An intuitive framework for real-time freeform modeling},
  author={Botsch, Mario and Kobbelt, Leif},
  journal={ACM Transactions on Graphics (TOG)},
  volume={23},
  number={3},
  pages={630--634},
  year={2004},
  publisher={ACM New York, NY, USA}
}

@inproceedings{jacobson2010mixed,
  title={Mixed finite elements for variational surface modeling},
  author={Jacobson, Alec and Tosun, Elif and Sorkine, Olga and Zorin, Denis},
  booktitle={Computer graphics forum},
  volume={29},
  number={5},
  pages={1565--1574},
  year={2010},
  organization={Wiley Online Library}
}

@phdthesis{tosun2008geometric,
  title={Geometric modeling using high-order derivatives},
  author={Tosun, Elif},
  year={2008},
  school={New York University}
}

@article{stein2018natural,
  title={Natural boundary conditions for smoothing in geometry processing},
  author={Stein, Oded and Grinspun, Eitan and Wardetzky, Max and Jacobson, Alec},
  journal={ACM Transactions on Graphics (TOG)},
  volume={37},
  number={2},
  pages={1--13},
  year={2018},
  publisher={ACM New York, NY, USA}
}

@inproceedings{jacobson2012smooth,
  title={Smooth shape-aware functions with controlled extrema},
  author={Jacobson, Alec and Weinkauf, Tino and Sorkine, Olga},
  booktitle={Computer Graphics Forum},
  volume={31},
  number={5},
  pages={1577--1586},
  year={2012},
  organization={Wiley Online Library}
}

@article{wang2021fast,
  title={Fast quasi-harmonic weights for geometric data interpolation},
  author={Wang, Yu and Solomon, Justin},
  journal={ACM Transactions on Graphics (TOG)},
  volume={40},
  number={4},
  pages={1--15},
  year={2021},
  publisher={ACM New York, NY, USA}
}

@inproceedings{thiery2018araplbs,
  title={{ARAPLBS}: Robust and efficient elasticity-based optimization of weights and skeleton joints for linear blend skinning with parametrized bones},
  author={Thiery, J-M and Eisemann, Elmar},
  booktitle={Computer Graphics Forum},
  volume={37},
  number={1},
  pages={32--44},
  year={2018},
  organization={Wiley Online Library}
}

@article{le2019direct,
  title={Direct delta mush skinning and variants},
  author={Le, Binh Huy and Lewis, JP},
  journal={ACM Trans. Graph.},
  volume={38},
  number={4},
  pages={113--1},
  year={2019}
}

@inproceedings{yang2021s3,
  title={S3: Neural shape, skeleton, and skinning fields for 3d human modeling},
  author={Yang, Ze and Wang, Shenlong and Manivasagam, Sivabalan and Huang, Zeng and Ma, Wei-Chiu and Yan, Xinchen and Yumer, Ersin and Urtasun, Raquel},
  booktitle={Proceedings of the IEEE/CVF Conference on Computer Vision and Pattern Recognition},
  pages={13284--13293},
  year={2021}
}

@article{liu2019neuroskinning,
  title={Neuroskinning: Automatic skin binding for production characters with deep graph networks},
  author={Liu, Lijuan and Zheng, Youyi and Tang, Di and Yuan, Yi and Fan, Changjie and Zhou, Kun},
  journal={ACM Transactions on Graphics (ToG)},
  volume={38},
  number={4},
  pages={1--12},
  year={2019},
  publisher={ACM New York, NY, USA}
}

@article{liao2023vinecs,
  title={{VINECS}: Video-based Neural Character Skinning},
  author={Liao, Zhouyingcheng and Golyanik, Vladislav and Habermann, Marc and Theobalt, Christian},
  journal={arXiv:2307.00842},
  year={2023}
}

@inproceedings{Chen2023INSR,
      title={Implicit Neural Spatial Representations for Time-dependent PDEs},
      author={Honglin Chen and Rundi Wu and Eitan Grinspun and Changxi Zheng and Peter Yichen Chen},
      booktitle={International Conference on Machine Learning},
      year={2023}
  }

@inproceedings{pan2021heterskinnet,
  title={{HeterSkinNet}: A Heterogeneous Network for Skin Weights Prediction},
  author={Pan, Xiaoyu and Huang, Jiancong and Mai, Jiaming and Wang, He and Li, Honglin and Su, Tongkui and Wang, Wenjun and Jin, Xiaogang},
  booktitle={Proceedings of the ACM on Computer Graphics and Interactive Techniques},
  volume={4},
  number={1},
  year={2021},
  organization={Association for Computing Machinery}
}

@article{ma2023tarig,
  title={{TARig}: Adaptive template-aware neural rigging for humanoid characters},
  author={Ma, Jing and Zhang, Dongliang},
  journal={Computers \& Graphics},
  year={2023},
  publisher={Elsevier}
}

@inproceedings{ouyang2020autoskin,
  title={{AutoSkin}: Skeleton-based Human Skinning with Deep Neural Networks},
  author={Ouyang, Xuming and Feng, Cunguang},
  booktitle={Journal of Physics: Conference Series},
  volume={1550},
  number={3},
  pages={032163},
  year={2020},
  organization={IOP Publishing}
}

@inproceedings{chen2021snarf,
  title={{SNARF}: Differentiable forward skinning for animating non-rigid neural implicit shapes},
  author={Chen, Xu and Zheng, Yufeng and Black, Michael J and Hilliges, Otmar and Geiger, Andreas},
  booktitle={Proceedings of the IEEE/CVF International Conference on Computer Vision},
  pages={11594--11604},
  year={2021}
}

@inproceedings{mosella2022skinningnet,
  title={{SkinningNet}: Two-stream graph convolutional neural network for skinning prediction of synthetic characters},
  author={Mosella-Montoro, Albert and Ruiz-Hidalgo, Javier},
  booktitle={Proceedings of the IEEE/CVF Conference on Computer Vision and Pattern Recognition},
  pages={18593--18602},
  year={2022}
}

@article{li2021learning,
  title={Learning skeletal articulations with neural blend shapes},
  author={Li, Peizhuo and Aberman, Kfir and Hanocka, Rana and Liu, Libin and Sorkine-Hornung, Olga and Chen, Baoquan},
  journal={ACM Transactions on Graphics (TOG)},
  volume={40},
  number={4},
  pages={1--15},
  year={2021},
  publisher={ACM New York, NY, USA}
}

@article{wu2020skinning,
  title={Skinning a parameterization of three-dimensional space for neural network cloth},
  author={Wu, Jane and Geng, Zhenglin and Zhou, Hui and Fedkiw, Ronald},
  journal={arXiv:2006.04874},
  year={2020}
}

@article{raissi2019physics,
  title={Physics-informed neural networks: A deep learning framework for solving forward and inverse problems involving nonlinear partial differential equations},
  author={Raissi, Maziar and Perdikaris, Paris and Karniadakis, George E},
  journal={Journal of Computational Physics},
  volume={378},
  pages={686--707},
  year={2019},
  publisher={Elsevier}
}

@article{xu2020rignet,
  title={{RigNet}: neural rigging for articulated characters},
  author={Xu, Zhan and Zhou, Yang and Kalogerakis, Evangelos and Landreth, Chris and Singh, Karan},
  journal={ACM Transactions on Graphics (TOG)},
  volume={39},
  number={4},
  pages={58--1},
  year={2020},
  publisher={ACM New York, NY, USA}
}

@article{kocabas2023hugs,
  title={{HUGS}: Human gaussian splats},
  author={Kocabas, Muhammed and Chang, Jen-Hao Rick and Gabriel, James and Tuzel, Oncel and Ranjan, Anurag},
  journal={arXiv:2311.17910},
  year={2023}
}

@inproceedings{ma2022neural,
  title={Neural Point-based Shape Modeling of Humans in Challenging Clothing},
  author={Ma, Qianli and Yang, Jinlong and Black, Michael J and Tang, Siyu},
  booktitle={International Conference on 3D Vision (3DV)},
  pages={679--689},
  year={2022},
  organization={IEEE}
}

@misc{warp2022,
title= {Warp: A High-performance Python Framework for GPU Simulation and Graphics},
author = {Miles Macklin},
month = {March},
year = {2022},
note= {NVIDIA GPU Technology Conference (GTC)},
howpublished = {\url{https://github.com/nvidia/warp}}
}

@misc{nanobind,
   author = {Wenzel Jakob},
   year = {2022},
   note = {https://github.com/wjakob/nanobind},
   title = {nanobind: tiny and efficient C++/Python bindings}
}

@inproceedings{kant2023invertible,
  title={Invertible Neural Skinning},
  author={Kant, Yash and Siarohin, Aliaksandr and Guler, Riza Alp and Chai, Menglei and Ren, Jian and Tulyakov, Sergey and Gilitschenski, Igor},
  booktitle={Proceedings of the IEEE/CVF Conference on Computer Vision and Pattern Recognition},
  pages={8715--8725},
  year={2023}
}

@inproceedings{wu2023magicpony,
  title={{MagicPony}: Learning articulated 3d animals in the wild},
  author={Wu, Shangzhe and Li, Ruining and Jakab, Tomas and Rupprecht, Christian and Vedaldi, Andrea},
  booktitle={Proceedings of the IEEE/CVF Conference on Computer Vision and Pattern Recognition},
  pages={8792--8802},
  year={2023}
}

@article{gropp2020implicit,
  title={Implicit geometric regularization for learning shapes},
  author={Gropp, Amos and Yariv, Lior and Haim, Niv and Atzmon, Matan and Lipman, Yaron},
  journal={Proceedings of the 37th International Conference on Machine Learning},
  year={2020}
}

@inproceedings{sitzmann2019siren,
author = {Sitzmann, Vincent
and Martel, Julien N.P.
and Bergman, Alexander W.
and Lindell, David B.
and Wetzstein, Gordon},
title = {Implicit Neural Representations
with Periodic Activation Functions},
booktitle = {Proceedings of the Conference on Neural Information Processing Systems},
year={2020}
}

@inproceedings{libigl,
author = {Jacobson, Alec and Panozzo, Daniele},
title = {libigl: prototyping geometry processing research in C++},
year = {2017},
isbn = {9781450354035},
publisher = {Association for Computing Machinery},
address = {New York, NY, USA},
url = {https://doi.org/10.1145/3134472.3134497},
doi = {10.1145/3134472.3134497},
booktitle = {SIGGRAPH Asia 2017 Courses},
articleno = {11},
numpages = {172},
location = {Bangkok, Thailand},
series = {SA '17}
}

@misc{gptoolbox,
  title = {{gptoolbox}: Geometry Processing Toolbox},
  author = {Alec Jacobson and others},
  note = {http://github.com/alecjacobson/gptoolbox},
  year = {2021},
}

@article{gingold1977smoothed,
  title={Smoothed particle hydrodynamics: theory and application to non-spherical stars},
  author={Gingold, Robert A and Monaghan, Joseph J},
  journal={Monthly Notices of the Royal Astronomical Society},
  volume={181},
  number={3},
  pages={375--389},
  year={1977},
  publisher={Oxford University Press Oxford, UK}
}

@article{lucy1977numerical,
  title={A numerical approach to the testing of the fission hypothesis},
  author={Lucy, Leon B},
  journal={Astronomical Journal},
  volume={82},
  pages={1013--1024},
  year={1977}
}

@article{sharp2020flipout, author = {Sharp, Nicholas and Crane, Keenan}, title = {You Can Find Geodesic Paths in Triangle Meshes by Just Flipping Edges}, journal = {ACM Trans. Graph.}, volume = {39}, number = {6}, year = {2020}, publisher = {ACM}, address = {New York, NY, USA}, }

@inproceedings{sharp2020laplacian, title={A laplacian for nonmanifold triangle meshes}, author={Sharp, Nicholas and Crane, Keenan}, booktitle={Computer Graphics Forum}, volume={39}, number={5}, pages={69--80}, year={2020}, organization={Wiley Online Library} }

@article{Sharp:2021:GPI, author = {Sharp, Nicholas and Gillespie, Mark and Crane, Keenan}, title = {Geometry Processing with Intrinsic Triangulations}, booktitle = {ACM SIGGRAPH 2021 courses}, series = {SIGGRAPH '21}, year = {2021}, publisher = {ACM}, address = {New York, NY, USA}, }

@article{gillespie2021integer,
    author = {Gillespie, Mark and Sharp, Nicholas and Crane, Keenan},
    title = {Integer Coordinates for Intrinsic Geometry Processing},
    journal = {ACM Trans. Graph.},
    volume = {40},
    number = {6},
    year = {2021},
    publisher = {ACM},
    address = {New York, NY, USA},
    url = {https://doi.org/10.1145/3478513.3480522},
    doi = {10.1145/3478513.3480522},
}

@article{Sawhney:2020:MCG,
author = {Sawhney, Rohan and Crane, Keenan},
title = {Monte Carlo Geometry Processing: A Grid-Free Approach to PDE-Based Methods on Volumetric Domains},
journal = {ACM Trans. Graph.},
volume = {39},
number = {4},
year = {2020},
publisher = {ACM},
address = {New York, NY, USA},
}

@article{Sawhney:2022:DND,
author = {Sawhney, Rohan and Seyb, Dario and Jarosz, Wojciech and Crane, Keenan},
title = {Grid-Free Monte Carlo for PDEs with Spatially Varying Coefficients},
journal = {ACM Trans. Graph.},
volume = {XX},
number = {X},
year = {2022},
publisher = {ACM},
address = {New York, NY, USA},
}

@article{Miller:2023:BVC,
author = {Miller, Bailey and Sawhney, Rohan and Crane, Keenan and Gkioulekas, Ioannis},
title = {Boundary Value Caching for Walk on Spheres},
journal = {ACM Trans. Graph.},
volume = {42},
number = {4},
year = {2023},
publisher = {ACM},
address = {New York, NY, USA},
}

@article{Sawhney:2023:WoSt,
author = {Sawhney, Rohan and Miller, Bailey and Gkioulekas, Ioannis and Crane, Keenan},
title = {Walk on Stars: A Grid-Free Monte Carlo Method for PDEs with Neumann Boundary Conditions},
journal = {ACM Trans. Graph.},
volume = {42},
number = {4},
year = {2023},
publisher = {ACM},
address = {New York, NY, USA},
}

@article{triwild,
author = {Hu, Yixin and Schneider, Teseo and Gao, Xifeng and Zhou, Qingnan and Jacobson, Alec and Zorin, Denis and Panozzo, Daniele},
title = {TriWild: robust triangulation with curve constraints},
year = {2019},
issue_date = {August 2019},
publisher = {Association for Computing Machinery},
address = {New York, NY, USA},
volume = {38},
number = {4},
issn = {0730-0301},
url = {https://doi.org/10.1145/3306346.3323011},
doi = {10.1145/3306346.3323011},
journal = {ACM Trans. Graph.},
month = {jul},
articleno = {52},
numpages = {15}
}

@misc{owl,
title={{OWL} - {A} {Productivity} {Library} for {OptiX}},
author={Wald, Ingo},
url={http://owl-project.github.io},
year={2020},
}

@article{rosales2019SurfaceBrush,
    author    = {Rosales, Enrique and Rodriguez, Jafet and Sheffer, Alla},
    title     = {SurfaceBrush: From Virtual Reality Drawings to Manifold Surfaces},
    journal   = {ACM Transaction on Graphics},
    year      = {2019},
    volume    = {38},
    number    = {4},
    doi       = {https://doi.org/10.1145/3306346.3322970},
    publisher = {ACM},
    address   = {New York, NY, USA}
}

@incollection{cgal:tet,
  author = {Jane Tournois and Noura Faraj and Jean-Marc Thiery and Tamy Boubekeur},
  title = {Tetrahedral Remeshing},
  publisher = {{CGAL Editorial Board}},
  edition = {{5.6.1}},
  booktitle = {{CGAL} User and Reference Manual},
  url = {https://doc.cgal.org/5.6.1/Manual/packages.html#PkgTetrahedralRemeshing},
  year = 2024
}

@inproceedings{dionne2013geodesic,
  title={Geodesic voxel binding for production character meshes},
  author={Dionne, Olivier and de Lasa, Martin},
  booktitle={Proceedings of the 12th ACM SIGGRAPH/Eurographics Symposium on Computer Animation},
  pages={173--180},
  year={2013}
}

@article{dionne2014geodesic,
  title={Geodesic binding for degenerate character geometry using sparse voxelization},
  author={Dionne, Olivier and de Lasa, Martin},
  journal={IEEE Transactions on Visualization and Computer Graphics},
  volume={20},
  number={10},
  pages={1367--1378},
  year={2014},
  publisher={IEEE}
}

@article{bang2018spline,
  title={Spline interface for intuitive skinning weight editing},
  author={Bang, Seungbae and Lee, Sung-Hee},
  journal={ACM Transactions on Graphics (TOG)},
  volume={37},
  number={5},
  pages={1--14},
  year={2018},
  publisher={ACM New York, NY, USA}
}

@article{xian2018efficient,
  title={Efficient $C^2$-weighting for image warping},
  author={Xian, Chuhua and Jin, Shuo and Wang, Charlie CL},
  journal={IEEE Computer Graphics and Applications},
  volume={38},
  number={1},
  pages={59--76},
  year={2018},
  publisher={IEEE}
}

@article{lu2021hard,
author = {Lu, Lu and Pestourie, Rapha\"{e}l and Yao, Wenjie and Wang, Zhicheng and Verdugo, Francesc and Johnson, Steven G.},
title = {Physics-Informed Neural Networks with Hard Constraints for Inverse Design},
journal = {SIAM Journal on Scientific Computing},
volume = {43},
number = {6},
pages = {B1105-B1132},
year = {2021},
doi = {10.1137/21M1397908},
URL = {https://doi.org/10.1137/21M1397908},
eprint = {https://doi.org/10.1137/21M1397908}
}

@inproceedings{zhong2023neural,
author = {Zhong, Fangcheng and Fogarty, Kyle and Hanji, Param and Wu, Tianhao and Sztrajman, Alejandro and Spielberg, Andrew and Tagliasacchi, Andrea and Bosilj, Petra and Oztireli, Cengiz},
title = {Neural fields with hard constraints of arbitrary differential order},
year = {2024},
publisher = {Curran Associates Inc.},
address = {Red Hook, NY, USA},
booktitle = {Proceedings of the 37th International Conference on Neural Information Processing Systems},
articleno = {992},
numpages = {26},
location = {New Orleans, LA, USA},
series = {NIPS '23}
}

@inproceedings{dugas2000softplus,
 author = {Dugas, Charles and Bengio, Yoshua and B\'{e}lisle, Fran\c{c}ois and Nadeau, Claude and Garcia, Ren\'{e}},
 booktitle = {Advances in Neural Information Processing Systems},
 editor = {T. Leen and T. Dietterich and V. Tresp},
 pages = {},
 publisher = {MIT Press},
 title = {Incorporating Second-Order Functional Knowledge for Better Option Pricing},
 url = {https://proceedings.neurips.cc/paper_files/paper/2000/file/44968aece94f667e4095002d140b5896-Paper.pdf},
 volume = {13},
 year = {2000}
}

\appendix

\section{Implementation Details}~\label{s:impl}

Our algorithm is primarily implemented in PyTorch~\cite{pytorch}, with the exception of closest-point queries in Section~\ref{ss:bc} and geometry-aware radius queries in Section~\ref{ss:kernel}.
For these, we provide our own low-level CUDA implementations.

For the geometry-aware kernel queries, we first build an SPH hash-grid in CUDA, with hash-grid cell sizes equal to the kernel radius.
For each query point, we loop over all of the points in the the neighboring cells.
If a point falls within the kernel radius, we trace a ray between it and the query point, as explained in Section~\ref{ss:kernel}.
For the ray-tracing itself, we rely on hardware-acceleration via the Optix \cite{Parker10OptiX} and OWL \cite{owl} libraries.
In practice, this means that the kernel queries are implemented as \emph{ray-generation} shaders in Optix parlance.
The closest point-queries rely on the Warp library \cite{warp2022}, which internally builds a bounding-volume hierarchy over the boundary mesh.

We make the CUDA kernels interoperable with PyTorch by relying on \texttt{nanobind}~\cite{nanobind}.
We use the fast generalized winding numbers implementation from \texttt{libigl}~\cite{libigl,barill2018fast}.

Before optimizing, we rescale meshes to fit into the $[0, 1]^d$ unit box.
We generate $\mathcal{X}$ by uniformly sampling the interior of the mesh, with an initial budget of $2^{11}$ points in $2$D and $2^{13}$ points in $3$D.
If a mesh of the interior is unavailable, we rely on fast winding number and perform rejection sampling until the target budget is reached.
Our initial learning rate is set to $0.2$.
Each iteration, we sample $M=2^{12}$ points in $2$D.
Since $\mathcal{X}$ is larger in $3$D compared to $2$D, using the same number of optimization samples would become prohibitively slow.
For this reason, we use fewer samples in $3$D, $M=2^{11}$.

We train for a total of $6000$ steps, and double the size of $\mathcal{X}$ a total of $3$ times at regular intervals throughout the optimization. Every time we upsample the representation, we lower the learning rate by a factor of $\frac{1}{3}$.

We found it important to lower the $\beta_2$ parameter of Adam~\cite{kingma2014adam} to $0.8$.
A higher value of $\beta_2$ results in the optimizer slowing down if the gradient changes direction.
This is useful in high-variance scenarios, but when optimizing variational energies, this is the opposite of what we want; in our scenario, gradients change directions often as boundary conditions propagate.

The sizes of the boundary condition parameters have been determined through our ablation study in Figure~\ref{fig:ablation}.
We set the value of the radii for the Neumann, Lagrange, and Dirichlet conditions equal to half the initial value of $\sigma$. We update the radius of the Lagrange and Dirichlet conditions as we upscale the point cloud.

\paragraph*{Evaluation on the Boundary.} When trying to evaluate the weights \emph{on} the boundary of a shape, the imprecision of the ray-tracing hardware can produce noisy results. Since it is highly unlikely that a random point will land close enough to the boundary for this to matter, this does not impact the quality of the weights. This means that our optimization remains unmodified. At evaluation time, we evaluate our weights \emph{on} the boundary using a few simple heuristics that work well in all of our tests.

To evaluate the kernel at a vertex $\vect{y}$, we compute the kernel weights between $\vect{y}$ and $\vect{x}\in\mathcal{X}$, but we trace rays between ${\vect{y} - 10^{-5}\vect{n}_{\vect{y}}}$ and $\vect{x}$. In other words, we nudge $\vect{y}$ slightly into the shape for the purposes of ray tracing alone.

In some cases, a vertex can be classified as outside of the shape by the generalized winding number, and/or completely occluded from any point $\vect{x}\in\mathcal{X}$ (e.g., a vertex slightly protrudes through a different triangle, common in VR ribbon drawings or triangle soups).
For all such vertices, we flood-fill their values from their neighbors using an averaging operation.
In the rare case that there are still any vertices with no weights, we run our kernel with no ray tracing.

\section{Smoothing Function}~\label{ap:funs}

Here we include definitions of the smoothing functions we rely on.
As defined by \citet{dugas2000softplus}, the \texttt{softplus} function is:
\begin{equation}
    \texttt{softplus}(t) = \log(1 + \exp(t)).
\end{equation}

\paragraph*{Lagrange Condition Mollifier}
In using $w$ to construct $\widetilde{\vect e}$ we must ensure that the partition of unity property \eqref{eq:partition} remains conserved, recognizing that the $\varepsilon$-neighborhoods of handles often overlap, e.g., because two bones share a vertex.
Denoting by $d_i(\DomainSample)$ the distance of a sample $\DomainSample$ to the $i$\textsuperscript{th} handle \emph{within} the $\varepsilon$-neighborhoods, we write:
\begin{equation}~\label{eq:lagrmollifier}
\begin{gathered}
    \widetilde e_i \left( \DomainSample \right) \coloneq
    \max_{j = 1 \ldots K} \left\{ w\left( d_j \left(\DomainSample\right) \right) \right\}
    \frac{
        w \left( d_i \left(\DomainSample\right) \right)
    }{
        \sum_{j=1}^K w\left( d_j \left(\DomainSample\right) \right)
    }.
\end{gathered}
\end{equation}
For a point $\DomainSample$ \emph{outside} of all $\varepsilon$-neighborhoods, we define $\widetilde e_i (\DomainSample) \coloneq 0$ to avoid division by zero.

In the case where $\DomainSample$ is within an $\varepsilon$-neighborhood of exactly one handle, \eqref{eq:lagrmollifier} simplifies to $\widetilde e_i \left( \DomainSample \right) = w(d_i(x))$.
If $\DomainSample$ happens to be inside multiple overlapping $\varepsilon$-neighborhoods, we use the bump function value of the closest handle and distribute it among all neighboring handles in proportion to the value of their respective bump function at $\DomainSample$. 
As a concrete example, if a point is equidistant to two different handles such that both of their bump functions equal $0.5$, both $\widetilde e_i$ will be assigned the value $0.25$, ensuring \eqref{eq:partition} still holds.

There are many possible \emph{bump functions} one could choose---for example, in all our experimets we use:
\begin{equation}
\begin{aligned}
    w'(t) &= \begin{cases}
        0 & t < 0 \\
        1 & t > 1 \\
        \frac{
            \exp(-\frac{1}{t})
        }{
            \exp(-\frac{1}{t}) + \exp(-\frac{1}{1 - t})
        } & \text{ otherwise,}
    \end{cases} \\
    w(t) &= 1 - w'\left(\frac{t^2}{\varepsilon^2}\right).
\end{aligned}
\end{equation}
Here, $w'$ has the convenient property that all of its higher-order derivatives disappear at $0$ and $1$; that way, the derivatives of $w'$ do not interfere with our energy computation when applying the chain rule.

\end{document}